\documentclass[superscriptaddress,secnumarabic,twocolumn,
 amssymb,amsmath,nobibnotes,aps,prd,showkeys,showpacs]{revtex4}
\usepackage{graphicx,epsfig,amssymb,amsmath,amstext}

\usepackage{graphicx}
\usepackage{bm}

\begin{document}

\title{Adjusting chaotic indicators to curved spacetimes}

\author{Georgios Lukes-Gerakopoulos}
\affiliation{Theoretical Physics Institute, University of Jena, 07743 Jena,
Germany}
\email{gglukes@gmail.com}

\begin{abstract}

 In this work, chaotic indicators, which have been established in the framework of
 classical mechanics, are reformulated in the framework of general relativity in
 such a way that they are invariant under coordinate transformation. For achieving
 this, the prescription for reformulating mLCE given by [Y.~Sota, S.~Suzuki, and
 K.-I.~Maeda, {\it Classical Quantum Gravity} {\bf 13}, 1241 (1996)] is adopted.
 Thus, the geodesic deviation vector approach is applied, and the proper time is
 utilized as measure of time. Following the aforementioned prescription, the chaotic
 indicators FLI, MEGNO, GALI, and APLE are reformulated. In fact, FLI has been
 reformulated by adapting other prescriptions in the past, but not by adapting the
 Sota et al.\ one. By using one of these previous reformulations of FLI, an
 approximative expression giving MEGNO as function of FLI has been applied on
 non-integrable curved spacetimes in a recent work. In the present work the
 reformulation of MEGNO is provided by adjusting the definition of the indicator to
 the Sota et al.\ prescription. GALI, and APLE are reformulated in the framework of
 general relativity for the first time. All the reformulated indicators by the
 Sota et al.\ prescription  are tested and compared for their efficiency to discern
 order from chaos.

\end{abstract}

\maketitle

\section{Introduction}\label{sec:Intro}

 The concept of a {\it chaotic} dynamical system is usually correlated with the
 property of a system exhibiting sensitive dependence on initial conditions (see, 
 e.g., the Devaney definition of chaos \cite{Devaney89}). Even though this
 correlation might be somehow misleading (see,  e.g., \cite{Banks94}), the
 sensitivity to initial conditions provides an efficient way to detect chaos.
 Therefore, various such detecting methods have been developed and established in
 the framework of classical celestial mechanics over the last decades (see,  e.g.,
 \cite{Contop02,Skokos10,Maffione}). 

 From this variety of methods we are going to investigate a group of indicators
 which use the evolution of deviation vectors along a given orbit. In the classical
 framework the deviation vector evolves in a space tangential to the phase space,
 the measure of this vector is taken to be Euclidean, and the time is an
 independent parameter. From the category of these indicators, the most renowned is
 the {\it maximal Lyapunov Characteristic Exponent} (mLCE) (see,  \cite{Skokos10}
 for a survey). Other similar indicators are: the {\it Fast Lyapunov Indicator}
 (FLI) \cite{Froesche97,Froesche05}, the {\it Mean Exponential Growth of Nearby 
 Orbits} (MEGNO) \cite{Cincotta00,Cincotta03}, the  {\it Generalized Alignment
 Index} (GALI) \cite{Skokos01,Skokos07}, the {\it Average Power Law Exponent} (APLE)
 \cite{LGVE08,LGVE09}. In classical mechanics the above mentioned indicators have
 been compared and studied for their efficiency several times (see,  e.g.,
 \cite{LGVE08,Maffione}).

 However, the definition and the efficiency of these indicators pose issues in the
 framework of General Relativity (GR) (see, e.g., \cite{Motter,Wu06} and references
 therein). Namely, one has to redefine the chaotic indicators in such a way that
 they will be invariant under coordinate transformations, and then to test these
 redefined indicators for their ability to detect chaos. In order to do the former,
 one has to find a way to define an invariant measure of the deviation vector in GR,
 and to choose an invariant time parameter. For the geodesic motion in curved
 spacetimes, which is the case we focus on, some suggestions to solve the above
 issues have already been provided. For instance, the indicators can be evaluated
 by applying the $3+1$ spacetime splitting approach \cite{Karas92} or by choosing
 the proper time as the time parameter \cite{Sota96} and using the invariant
 measure of the deviation vector either derived by the geodesic deviation equations
 \cite{Sota96} or by the two nearby orbits approximation \cite{Wu06}. In this
 study, the guideline of Sota et al.\ \cite{Sota96} was preferred for adjusting the
 chaotic indicators to the GR framework. However, if we depart from the geodesic
 motion, for example by taking into account the spin of the test particle, then
 approaches stemming from the $3+1$ splitting \cite{Karas92} are maybe preferable
 for addressing the aforementioned issues (see, e.g., \cite{Hartl03,Han08}).

 On the other hand, the above indicators are not the only methods which have been
 employed for detecting chaos in relativistic systems. Frequency analysis
 techniques, which were applied initially in the framework of classical mechanics
 (see, e.g., \cite{Laskar93}), have been lately applied in the GR framework as well
 (see, e.g., \cite{SemSuk,LG12}); the same holds for the recurrence analysis
 techniques (see,  \cite{Marwan07} for a review) which were also applied recently
 in curved spacetimes (see, e.g., \cite{Kopac,SemSuk}). Both frequency analysis and
 recurrence analysis techniques are applied on time series, which makes them 
 appropriate for the observational data post-analysis. Moreover, the recurrence
 analysis is able to discern deterministic chaos from stochastic noise, which might
 be very useful when the signal is embedded in noise. 
 
 Yet another kind of approach are the basin boundaries \cite{basin}, which take
 advantage of the fractal geometry of a non-integrable system to detect the 
 existence of chaos. The methods which use the curvature of a spacetime to search
 for chaos \cite{Sota96,Szyd} are also Geometrical.
 
 The background spacetime of a rapidly spinning neutron star suggested in \cite{MSM}
 provides the non-integrable dynamical system to test the adjusted chaotic
 indicators. We are going to refer to this spacetime as Manko, Sanabria-G{\'o}mez,
 Manko or briefly MSM from now on. MSM background belongs to a broader family of
 spacetimes describing the surrounding spacetime of neutron stars; this family of
 spacetimes was introduced in \cite{Manko95} and revisited in \cite{Pappas13}, and
 their astrophysical importance was investigated in \cite{Pappas13,Berti04}. Now,
 from the dynamical point of view, since the existence of chaos in MSM background
 has already been revealed in \cite{Dubeibe07,Han08b,Seyrich12}, the MSM spacetime 
 provides the appropriate background for testing chaotic indicators on geodesic
 orbits.

 The integration scheme applied to evolve these geodesic orbits along with the 
 geodesic deviation equations is a symmetric, reversible integrator called
 integrator for geodesic equations of motion (IGEM) \cite{Seyrich12}. IGEM has been
 designed to evolve strongly chaotic orbits efficiently and to preserve the
 constants of motion. IGEM has been tested and compared with other integrators 
 in the MSM spacetime \cite{Seyrich12}. From the above comparison IGEM appears to
 be the most appropriate for the present study.
  
 The paper is organized as follows. Sec.~\ref{sec:MSM} provides a brief description
 of the curved spacetime in which the chaotic indicators are tested. A brief survey
 on the geodesic and geodesic deviation equation of motion follows in
 Sec.~\ref{sec:devges}. The chaotic indicators and their invariant reformulation in
 curved spacetimes are presented in Sec.~\ref{sec:Indices}. Numerical examples
 of these indicators are given in Sec.~\ref{sec:NumEx}. Sec.~\ref{sec:Conc}
 surveys the main results, and in Appendix~\ref{sec:NumAc} the accuracy of the
 integrating scheme is discussed.

 \section{The Manko, Sanabria-G{\'o}mez, Manko spacetime}\label{sec:MSM}

 It has been already mentioned that MSM belongs to a family of spacetimes that
 were designed to model neutron stars (see,  e.g., \cite{Pappas13,Berti04}). The MSM
 spacetime is asymptotically flat, axisymmetric and stationary; it describes the
 ``exterior field of a charged, magnetized, spinning deformed mass'' \cite{MSM}.
 The MSM is a five-parameter vacuum solution, it depends on the mass $m$, the spin
 per unit mass $a$, the total charge $q$, the magnetic dipole moment ${\cal M}$,
 and the mass-quadrupole moment ${\cal Q}$. However, the two  latter quantities are
 functions of the first three real parameters and of two other real parameters,
 i.e., $\mu$ and $b$,
 \begin{eqnarray} \label{eq:MagQuad}
  {\cal M} &=& \mu+q (a-b)~~, \nonumber \\
  {\cal Q} &=& -m(d-\delta-a~b+a^2)~~, 
 \end{eqnarray}
 where
 \begin{eqnarray} \label{eq:deltad}
   \delta &:=& \frac{\mu^2-m^2 b^2}{m^2-(a-b)^2-q^2}~~, \nonumber \\
   d &:=& \frac{1}{4}[m^2-(a-b)^2-q^2]~~.  
 \end{eqnarray}
   
 The Weyl-Papapetrou line element of the MSM spacetime in prolate spheroidal
 coordinates $t,~x,~\phi,~y$ is
 \begin{equation}\label{eq:WPcyl}
  ds^2=g_{tt}~dt^2+g_{t\phi}dt~d\phi+g_{\phi\phi}+g_{xx}d\rho^2
 +g_{yy}dz^2~~,\\ 
 \end{equation}
 where
 \begin{eqnarray}\label{eq:MSMmetric}
  g_{tt} & = & -f~~, \nonumber \\
  g_{t\phi} & = & f \omega~~, \nonumber \\
  g_{\phi\phi} & = & \frac{k^2(x^2-1)(1-y^2)}{f}-f \omega^2 ~~, \\
  g_{xx} & = &  \frac{k^2 e^{2\gamma}}{f}\frac{x^2-y^2}{x^2-1}~~. \nonumber \\
  g_{yy} & = &  \frac{k^2 e^{2\gamma}}{f}\frac{x^2-y^2}{1-y^2}~~. \nonumber 
 \end{eqnarray} 
 
 The functions $f$,~$\omega$,~and~$\gamma$ are
 \begin{eqnarray} \label{eq:MetrFunc}
  f &=& C/D~~, \nonumber \\
  e^{2\gamma} &=& C/16 k^8 (x^2-y^2)^4~~,\\
  \omega &=& (y^2-1)F/C~~,\nonumber
 \end{eqnarray}
 and
 \begin{equation} \label{eq:kappa}
  k :=\sqrt{d+\delta}~~. 
 \end{equation}

  The functions $C$,~$D$,~and~$F$ are
 \begin{eqnarray} \label{eq:EDF}
  C &=& R^2+\lambda_1\lambda_2 S^2~~, \nonumber \\
  D &=& E+R P+\lambda_2 S T~~, \\
  F &=& R T-\lambda_1 S P~~,\nonumber
 \end{eqnarray}
 where
 \begin{equation} \label{eq:lambda}
  \lambda_1=k^2 (x^2-1),~~\lambda_2=y^2-1~~.
 \end{equation}

 The functions $P$,~$R$,~$S$~and~$T$ are 
 \begin{eqnarray}\label{eq:PRST}
  P & := & 2 \{k m x [(2 k x+m)^2-2 y^2 (2 \delta +a b -b^2)\nonumber \\
    & - & a^2+b^2 -q^2]-2 k^2 q^2 x^2-2 y^2 (4 \delta d-m^2 b^2)\}~~, 
   \nonumber \\
  R & := & 4 [k^2 (x^2-1)+\delta (1-y^2)]^2 \nonumber \\
    &  + & (a-b) [(a-b)(d-\delta)-m^2 b +q~\mu](1-y^2)^2~~,\nonumber \\
    & ~ &  \\
  S & := & -4 {(a-b)[k^2(x^2-y^2)+2 \delta y^2]+y^2 (m^2 b-q~\mu)}~~,
  \nonumber \\
  T & := & 4(2 k m b x+2 m^2 b-q~\mu)[k^2 (x^2-1)+\delta (1-y^2)]
  \nonumber \\
    & + & (1-y^2)\{(a-b)(m^2 b^2 -4 \delta d)  \nonumber \\
    & - & (4 k m x+2 m^2-q^2) [(a-b)(d-\delta)-m^2 b +q~\mu]\}. \nonumber 
 \end{eqnarray}

 It is useful to mention that in the numerical calculations it is better to use the
 following combinations and expressions, in order to avoid numerical errors when 
 the orbits approach the static limit $g_{tt}=C=0$,
 \begin{eqnarray}
  \left[\frac{e^{2\gamma}}{f}\right] &=& \frac{D}{16k^8(x^2-y^2)^4}~~, \nonumber \\
  \left[f \omega \right] &=& \lambda_2 \frac{F}{D}~~, \nonumber
 \end{eqnarray}
 and 
 \begin{eqnarray}
  g_{\phi\phi} &=& 
   -\frac{\lambda_2}{D}\left[\lambda_1\left(C+2(RP+\lambda_2 ST)\right)
  +\lambda_1 P^2+ \lambda_2 T^2\right] \nonumber \\
  &=& -\left[2\lambda_1\lambda_2+\frac{\lambda_2}{D}(\lambda_1(P^2-E)
 +\lambda_2 T^2)\right]~~.\nonumber
 \end{eqnarray} 

 The numerical calculations were done in prolate spheroidal coordinates $x,~y$,
 but the results are presented in cylindrical coordinates $\rho,~z$ to facilitate
 the comparison with previous works \cite{Dubeibe07,Han08b,Seyrich12}. The two
 coordinate systems relate through the transformation
 \begin{equation} \label{eq:TrCylPS}
  \rho=k \sqrt{(x^2-1)(1-y^2)},~~ z=k x y~~.
 \end{equation} 

 \section{Geodesic and geodesic deviation}\label{sec:devges}

 The fact that geodesic motion in MSM background exhibits chaotic behavior was
 shown in \cite{Dubeibe07,Han08b,Seyrich12} mainly by studying Poincar\'{e}
 sections, but also by applying the FLI indicator \cite{Han08} as defined in
 \cite{Wu06}, and by the means of frequency analysis \cite{Seyrich12}.

 For finding Poincar\'{e} sections, we have to evolve the geodesic equations
 \begin{equation} \label{eq:geod}
  \ddot{x}^\alpha+\Gamma^\alpha_{\beta\gamma}\dot{x}^\beta\dot{x}^\gamma=0~~,
 \end{equation}
 where the dot corresponds to a derivative with respect to the proper time $\tau$,
 and $\Gamma^\alpha_{\beta\gamma}$ are the Christoffel symbols. The greek indices
 correspond to the whole spacetime. 

 The geodesic equations (\ref{eq:geod}) are the Euler-Lagrange equations of the
 Lagrangian function
 \begin{equation} \label{eq:Lagr}
  L=\frac{1}{2}g_{\alpha\beta} \dot{x}^\alpha \dot{x}^\beta~~,
 \end{equation}
 which is a constant of motion $L=-1$, and expresses the conservation of the four
 velocity of the test particle. The stationarity of the MSM spacetime provides the
 second constant
 \begin{equation} \label{eq:Energy}
  p_t=\frac{L}{\dot{t}}=-E~~,
 \end{equation}
 which is the energy of the test particle, while the axisymmetry provides the
 third constant
 \begin{equation} \label{eq:AnMomZ}
  p_\phi=\frac{L}{\dot{\phi}}=L_z~~,
 \end{equation}
 which is the azimuthal component of the test particle's angular momentum. By the
 last two constants the system is reduced to two degrees of freedom, and therefore,
 the Poincar\'{e} section can be used for detecting chaos in MSM spacetime
 backgrounds.

 However, for including indicators in the study depending on deviation vectors
 like FLI, we need the geodesic deviation equations
 \begin{equation} \label{eq:GeoDev}
  \ddot{\xi}^\alpha+2\Gamma^\alpha_{\beta\gamma}\dot{x}^\beta\dot{\xi}^\gamma+
 \frac{\partial \Gamma^\alpha_{\beta\gamma}}{\partial x^\delta}
 \dot{x}^\beta\dot{x}^\gamma\xi^\delta=0~~,
 \end{equation}
 which show how two initially nearby geodesic orbits ``$x^\alpha$'' and
 ``$x^\alpha+\xi^\alpha$'' diverge from each other. $\xi^\alpha$ is the deviation
 vector, whose behavior plays a major role in distinguishing order from chaos as
 discussed in the next section.  

 \section{Chaotic indicators} \label{sec:Indices}

 The measure of the deviation vector for a regular orbit grows linearly, while for
 a chaotic the growth is exponential or at least it follows a power law (see, e.g.,
 \cite{LGVE08}). This fact is the characteristic which mLCE, FLI, MEGNO, and APLE
 are designed to track. In order to have an invariant measure of the deviation
 vector in the phase space, Sota et al.\ \cite{Sota96} defined the quantity
 \begin{equation} \label{eq:SotaDis}
   \Xi^2=g_{\alpha\beta}\xi^\alpha \xi^\beta+g_{\alpha\beta}
 \frac{D\xi^\alpha}{d\tau}\frac{D\xi^\beta}{d\tau}~~,
 \end{equation}
 where the covariant derivative
 \begin{equation}\label{eq:CovDev}
   \frac{D\xi^\alpha}{d\tau}=\dot{\xi}^\alpha
   +\Gamma^\alpha_{\beta\gamma} \dot{x}^\beta \xi^\gamma
 \end{equation}
 provides the divergence of the velocities. 

 In order to ensure that $\Xi^2$ stays positive throughout the simultaneous
 evolution of the Eqs.~(\ref{eq:geod}),~(\ref{eq:GeoDev}), we have to ensure that
 $\xi^\alpha$ and $\displaystyle \frac{D\xi^\alpha}{d\tau}$ will remain spacelike.
 The prescription for this \cite{Sota96,Wu06} is to choose initial conditions for
 $\xi^\alpha,~\displaystyle\frac{D\xi^\alpha}{d\tau}$ such that
 \begin{equation} \label{eq:InDevPre}
  \xi^\alpha \dot{x}_\alpha= \frac{D \xi^\alpha}{d \tau} \dot{x}_\alpha=0~~.
 \end{equation}
 However, condition (\ref{eq:InDevPre}) is not the only way to ensure $\Xi^2>0$,
 and in \cite{Wu06} other options are discussed. Anyway, for the numerical
 calculations done in this study the initial prescription of Sota et al.\
 \cite{Sota96} is followed.  

 To address the issue of invariant time measure, whenever the definition of an
 indicator asks for a time parameter, the proper time is utilized. This parameter
 should be normalized by a typical time scale, e.g., $\tau_{ts}\approx G m/c^3$
 \cite{Sota96}. Throughout the article, geometric units are used, i.e., G=c=1, and
 the value of the mass $m$ of the central object is chosen to be of order of one,
 thus for simplicity, and without loss of generality, this time scale is set to
 be $\tau_{ts}=1$.

 \subsection{mLCE}\label{subsec:mLCE}

 The maximal Lyapunov Characteristic Exponent 
 \begin{equation}\label{eq:mLCE}
  \displaystyle \textrm{mLCE}=\lim_{\tau\rightarrow\infty}\frac{1}{\tau}
  \ln{\frac{\Xi(\tau)}{\Xi(0)}}
 \end{equation}
 is the most renowned chaotic indicator (see,  \cite{Skokos10} for a review). The
 limit at  infinity makes mLCE unrealistic for numerical studies, and the finite 
 form of mLCE 
 \begin{equation}\label{eq:FmLCE}
  \displaystyle \textrm{FmLCE}=\frac{1}{\tau} \ln{\frac{\Xi(\tau)}{\Xi(0)}}
 \end{equation} 
 is used instead. In Eq.~(\ref{eq:FmLCE}) $\tau$ is sufficiently large. However, in
 the literature FmLCE is usually referred to as mLCE, which is adopted also in
 this article. Several techniques to find the invariant form of mLCE have already
 been suggested for geodesic flow in curved space, and a survey of these techniques
 can be found in \cite{Wu06}. 

 One category of these techniques uses a ``shadow'' orbit instead of evolving the 
 geodesic deviation Eqs.~(\ref{eq:GeoDev}). This shadow orbit is a geodesic orbit
 with initial conditions very near to the orbit under study, and the distance
 in the configuration space between these two orbits is used instead of $\Xi$. The
 shadow technique provides probably an easier way to discover whether an orbit is
 chaotic or not than the geodesic deviation technique does, because one just has
 to evolve two nearby orbits by computing the geodesic Eqs.~(\ref{eq:geod}).
 However, since the evolution of two orbits in a curved spacetime is not as exact
 as evolving the geodesic deviation Eqs.~(\ref{eq:GeoDev}) in a spacetime tangent
 to the phase space where the orbital motion takes place, this approximation has a
 cost. Namely, even if we get a value of the mLCE near to the real mLCE (see,  e.g.,
 the numerical examples in \cite{Wu06}), we lose the invariance of the mLCE 
 indicator by using the shadow approximation. 

 The category of techniques using geodesic deviation equations splits into two
 subcategories. One subcategory uses the definition of $\Xi$ given by Sota et al.\
 \cite{Sota96} (Eq.~(\ref{eq:SotaDis})) and the other measures the distance $\Xi$
 only in the configuration space i.e., $\Xi^2=g_{\alpha\beta}\xi^\alpha \xi^\beta$.
 Now, the fact that the latter subcategory confines itself to a subspace of the
 space tangent to the phase space raises the question whether this technique can
 indeed find the invariant value of mLCE or it just distinguishes order from chaos,
 which would mean that this subcategory shares the same drawback with the technique
 of shadow orbits. On the other hand, the subcategory using the measure of $\Xi$
 as given in Eq.~(\ref{eq:SotaDis}) does not suffer from such ambiguity, since
 $\Xi$ is defined in the phase space. The latter has been in fact used for the
 reformulation of mLCE in \cite{Sota96}.

 The principle of chaos detection behind the mLCE indicator is the following. When
 an orbit is regular, which means that on average $\Xi(\tau)$ grows linearly, then
 from Eq.~(\ref{eq:FmLCE}) it is easy to show that
 $$\displaystyle \textrm{mLCE} \propto \frac{\ln{\tau}}{\tau}~~.$$ Thus, for a
 regular orbit, $$\lim_{\tau \rightarrow \infty} \textrm{mLCE}\rightarrow 0~~.$$
 When an orbit is chaotic, which means usually that $\Xi(\tau)$ grows exponentially,
 e.g., $\Xi(\tau) \propto e^{\lambda \tau}$ where $\lambda$ is constant, then from
 Eq.~(\ref{eq:FmLCE}) one gets $$\displaystyle \textrm{mLCE} \propto \lambda~~.$$
 Thus, for a chaotic orbit,
 $$\lim_{\tau \rightarrow \infty} \textrm{mLCE}\rightarrow \lambda~~.$$ 

 \subsection{FLI} \label{subsec:FLI}
     
 The principles behind the mLCE indicator hold also for the Fast Lyapunov Indicator
 \cite{Froesche97,Froesche05},
  \begin{equation}\label{eq:FLI}
  \displaystyle \textrm{FLI}=\ln{\frac{\Xi(\tau)}{\Xi(0)}}~~.
 \end{equation} 
 The difference here is that in order to discern a chaotic orbit from a regular
 orbit, one has to define a time dependent limit. This limit depends on the maximum
 value of the FLI ($\textrm{FLI}_{max}$) that a regular orbit reaches at a given
 time. Then the $\textrm{FLI}_{max}$ value is compared with the FLI value reached
 by the other orbits at this given time. If the FLI value of an orbit is above
 $\textrm{FLI}_{max}$, then the orbit is characterized as chaotic. In fact, usually
 this limit is set as $\textrm{FLI}_{max}$ plus a relatively arbitrary ``safety''
 value. For instance, if the maximum FLI value is $\textrm{FLI}_{max}$ in the
 finite proper time $\tau_f$, then the limit can be set to
 $\textrm{FLI}_0=\textrm{FLI}_{max}+\textrm{Constant}$, and any orbit whose
 $\textrm{FLI}>\textrm{FLI}_0$ is characterized as chaotic. For a detailed
 discussion on the $\textrm{FLI}_0$ issue refer to Sec.~3.2 of \cite{LGVE08}. 

 In the framework of general relativity a reformulation of FLI was proposed in
 \cite{Wu06} by employing the shadow orbit technique already discussed in
 Sec.~\ref{subsec:mLCE}. By the means of this approximative technique, FLI has
 already been applied in a few works (see,  e.g., \cite{Han08b,SemSuk}), but FLI has
 not yet been tested by applying the geodesic deviation technique according to
 the author's knowledge. 

 \subsection{MEGNO} \label{subsec:MEGNO}

 The basic definition of the Mean Exponential Growth of Nearby Orbits
 \cite{Cincotta00,Cincotta03} is    
  \begin{equation}\label{eq:MEGNO}
  \displaystyle
  \textrm{MEGNO}(\tau_f)=
  \frac{2}{\tau_f}\int_0^{\tau_f}\frac{\dot{\Xi}}{\Xi}\tau d\tau~~,
 \end{equation} 
 where $\tau_f$ is the finite proper time until which the equations of motion
 (\ref{eq:geod}), (\ref{eq:GeoDev}) are computed. A quite good approximation for
 MEGNO correlates it with the FLI indicator \cite{Mestre11}, i.e.,
 \begin{equation}\label{eq:ApMEGNO}
  \displaystyle
  \textrm{MEGNO}(\tau_f)=2\left[\textrm{FLI}(\tau_f)-<\textrm{FLI}(\tau_f)>\right]~~,
 \end{equation}
 where $<\textrm{FLI}(\tau_f)>$ is the mean value of $\textrm{FLI}$ until the time
 $\tau_f$. However, MEGNO defined in the form (\ref{eq:MEGNO}) suffers from big
 value oscillations; for this reason, the average value of MEGNO,
  \begin{equation}\label{eq:AvMEGNO}
  \displaystyle
  <\textrm{MEGNO}(\tau_f)>=
  \frac{1}{\tau_f}\int_0^{\tau_f} MEGNO(\tau) d\tau~~,
 \end{equation} 
 is more useful. In fact, from now on we are going to refer to average MEGNO simply
 as MEGNO. The advantage of MEGNO over FLI is that it has a time independent limit
 by which an orbit is characterized as a chaotic or a regular one. For a regular
 orbit, MEGNO tends asymptotically to two, while if the orbit is chaotic it tends
 asymptotically to infinity.  

 Recently, in the last article of the \cite{SemSuk} series, the MEGNO was tested in
 curved spacetimes describing a Schwarzschild black hole surrounded by a thin disc
 or a ring. The authors of this article used the approximation given by
 Eq.~(\ref{eq:ApMEGNO}), and applied the shadow orbit technique to approximate the
 deviation vector. In the present study, another approach is followed. By using the
 approximation
 $$\displaystyle \dot{\Xi}(\tau)=\frac{\Xi(\tau)-\Xi(\tau-d\tau)}{d\tau}~~,$$
 and by rewriting the formula (\ref{eq:MEGNO}) in discrete form, we arrive at
  \begin{equation}\label{eq:DisMEGNO}
  \displaystyle
  \textrm{MEGNO}(\tau_f)=
  \frac{2}{\tau_f}\sum_{i=0}^{N}
 \left(1-\frac{\Xi(\tau_i-d\tau_i)}{\Xi(\tau_i)}\right)\tau_i~~,
 \end{equation} 
 where $\displaystyle \tau_f=\sum_{i=0}^{N}\tau_i$. Respectively, the discrete form
 of Eq.~(\ref{eq:AvMEGNO}) is
  \begin{equation}\label{eq:DisAvMEGNO}
  \displaystyle
  <\textrm{MEGNO}(\tau_f)>=
  \frac{1}{\tau_f}\sum_{i=0}^{N} MEGNO(\tau_i) d\tau_i~~,
 \end{equation} 
 where $d\tau_i=\tau_i-\tau_{i-1}$ is practically the integration step used in the
 numerical calculations.

 \subsection{APLE} \label{subsec:APLE}

 The Average Power Law Exponent \cite{LGVE08,LGVE09},
 \begin{equation}\label{eq:APLE}
  \displaystyle \textrm{APLE}=
  \lim_{\tau\rightarrow\infty}\frac{\ln{\frac{\Xi(\tau)}{\Xi(0)}}}{\ln{\tau}}~~,  
 \end{equation}
 was defined in order to detect ``metastable'' behaviors of weakly chaotic orbits.
 During this ``metastable'' phase the measure of the deviation vector increases
 following nearly a power law $\Xi(\tau)\propto\tau^{p}$. This phase ends when the
 measure of the vector begins to grow exponentially. Like the MEGNO, APLE has a
 limit to which regular orbits converge; this limit is the value one. If the orbit 
 is weakly chaotic, then APLE will oscillate around a value equal to $p$ during the 
 metastable phase. After the metastable phase, or if the orbit is strongly chaotic,
 the value of APLE goes to infinity following the exponential growth of the
 deviation vector.

 In order to avoid a nullification of the denominator in Eq.~(\ref{eq:APLE}),
 we can use various numerical tricks, which do not compromise the efficiency of
 the indicator (see,  \cite{LGVE08} for a detailed discussion). For the purpose of
 this work, $\ln{(1+\tau)}$ was utilized; thus,
 \begin{equation}\label{eq:APLEN}
  \displaystyle \textrm{APLE}=
  \lim_{\tau\rightarrow\infty}\frac{\ln{\frac{\Xi(\tau)}{\Xi(0)}}}{\ln{(1+\tau)}}  
 \end{equation}
 is used in the numerical examples of Sec.~\ref{sec:NumEx} instead of the
 Eq.~(\ref{eq:APLE}). For $\tau>>1$, definition (\ref{eq:APLE}) is numerically 
 equivalent to formula (\ref{eq:APLEN}).

 \subsection{GALI} \label{subsec:GALI} 

 The Generalized Alignment Index \cite{Skokos07} is a generalization of the Smaller
 Alignment Index (SALI) \cite{Skokos01} (also called Alignment Index
 \cite{Voglis02}). GALI differs from the indicators discussed above because it does
 not depend on the rate by which a deviation vector grows, but on whether two or
 more deviation vectors with different initial directions will get aligned or not.
 GALI is identical to SALI when only two deviation vectors are used. 

 In particular, GALI uses the following properties of the deviation vectors. In the
 case of a chaotic orbit, two or more deviation vectors with different and arbitrary
 initial orientations will become parallel or anti-parallel exponentially fast. The
 speed by which this will happen depends on the value of the mLCE. On the other
 hand, in the case of regular motion, an orbit moves on a torus, and two or more
 deviation vectors with different and arbitrary initial orientations will become
 tangent to that torus with time. However, if the torus is $N$-dimensional, where
 $N\geq 2$, the orientation of the deviation vectors will remain, in general,
 different. If the torus is one-dimensional then the deviation vectors will become
 parallel or anti-parallel, but the time will follow a power law. Initially these
 properties have been investigated for the spectral distance techniques (see,  e.g.,
 \cite{VCE}), but by the introduction of SALI \cite{Skokos01} a more simple and
 efficient technique to detect chaos has been provided. 

 In \cite{Skokos01} SALI was defined as
 \begin{equation} \label{eq:C1SALI}
  \textrm{SALI}=\min{\left\{\left|\mathbf{w}-\mathbf{w}'\right|,
 \left|\mathbf{w}+\mathbf{w}'\right|\right\}}~~,
 \end{equation}
 where $\mathbf{w}$ and $\mathbf{w}'$ are the deviation vectors of classical
 mechanics normalized to unity by their Euclidean norm. Another way to define SALI
 is to take the cross product of these vectors, .i.e.,
 \begin{equation}\label{eq:C2SALI}
  \textrm{SALI}=\left|\mathbf{w}\times \mathbf{w}'\right|=\sin{ \theta}~~,
 \end{equation}
 where $\theta$ is the angle between the two vectors. The definition
 (\ref{eq:C2SALI}) reveals that SALI, in fact, measures the surface defined by the
 two vectors. 
 
 In the case of a system with two degrees of freedom or more, SALI goes to zero for 
 a chaotic orbit, while for a regular orbit it remains non-zero. In the case of a
 two dimensional map, SALI always goes to zero, but for chaotic orbits this happens 
 exponentially fast, while for regular orbits $\textrm{SALI}\propto t^{-q}$, where 
 $q \approx 2 $. This kind of power laws, in fact, provide the means for GALI to
 find the dimension of a torus in multidimensional systems \cite{Skokos07}. The
 advantage of GALI over the other indicators is exactly this ability, but in order
 to use it, we have to evolve more than two deviation vectors. Thus, the advantage
 of GALI comes with a certain computational cost.

 In curved spacetimes the Euclidean norm is not invariant under coordinate
 transformations, thus we cannot normalize the generalized deviation vector
 (Eq.~(\ref{eq:SotaDis})) defined by $\xi^\alpha$ and
 $\displaystyle \frac{D \xi^\alpha}{d \tau}$ to unity. This certainly is a problem
 for the definition (\ref{eq:C1SALI}), because for parallel vectors
 $\left|\mathbf{w}-\mathbf{w}'\right|$ will not go to zero and for anti-parallel
 vectors $\left|\mathbf{w}+\mathbf{w}'\right|$ will not go to zero.

 On the other hand, in the definition (\ref{eq:C2SALI}) we really don't depend on
 the strict normalization of the deviation vector to unity, the only thing we need
 is to limit the growth of the components $\xi^\alpha$ and
 $\displaystyle \frac{D \xi^\alpha}{d \tau}$. In order to do that we can divide
 them by the measures of the corresponding vectors, i.e.,
 $\displaystyle \frac{\xi^\alpha}{\sqrt{\xi_\kappa \xi^\kappa}}$ and $\displaystyle
 \frac{\frac{D \xi^\alpha}{d \tau}}{\sqrt{\frac{D \xi_\kappa}{d \tau}
 \frac{D \xi^\kappa}{d \tau}}}$. Then we can use the outer products of one pair 
 $\xi^\alpha$, $\zeta^\alpha$ of the deviation vectors and their corresponding
 velocities $\displaystyle \frac{D \xi^\alpha}{d \tau}$
 $\displaystyle \frac{D \zeta^\alpha}{d \tau}$ to provide a similar definition of
 SALI to Eq.~(\ref{eq:C2SALI}), i.e.,
 \begin{eqnarray} \label{eq:OuterProd}
  OI_{\alpha\beta} &=& \eta_{\alpha\beta\gamma\delta} 
 \frac{\xi^\gamma}{\sqrt{\xi_\kappa\xi^\kappa}}
 \frac{\zeta^\delta}{\sqrt{\zeta_\nu\zeta^\nu}}~~, \\
  OII_{\alpha\beta} &=& \eta_{\alpha\beta\gamma\delta} 
 \frac{\frac{D \xi^\gamma}{d \tau}}{\sqrt{\frac{D \xi_\kappa}{d \tau}\frac{D \xi^\kappa}{d \tau}}}
 \frac{\frac{D \zeta^\delta}{d \tau}}{\sqrt{\frac{D \zeta_\nu}{d \tau}\frac{D \zeta^\nu}{d \tau}}}~~,
 \end{eqnarray}
 where $\eta_{\alpha\beta\gamma\delta}$ is the Levi-Civita density tensor 
 \begin{equation} \label{eq:LevCivD}
  \eta_{\alpha\beta\gamma\delta}=\sqrt{-g}~\epsilon_{\alpha\beta\gamma\delta}~~,
 \end{equation}
 and $\epsilon_{\alpha\beta\gamma\delta}$ is the Levi-Civita symbol with
 $\epsilon_{0123}=-1$.

 If the deviation vectors $\xi^\alpha$, $\zeta^\alpha$ and their velocities
 $\displaystyle \frac{D \xi^\alpha}{d \tau}$, 
 $\displaystyle \frac{D \zeta^\alpha}{d \tau}$ are parallel, then both
 $OI_{\alpha\beta}$ and $OII_{\alpha\beta}$ are null. Thus, we can define the
 quantity
 \begin{equation} \label{eq:GRSALI}
  SALI=\sum_{\alpha=0}^3\sum_{\beta=0}^3
 \left(OI_{\alpha\beta}+OII_{\alpha\beta}\right)~~,
 \end{equation}
 which will go to zero for chaotic orbits and remain non-zero for regular orbits.
 In order to define GALI, we can use outer products for multiple deviation vectors
 and their corresponding velocities similar to (\ref{eq:OuterProd}), and sum these 
 outer products as has been suggested for SALI in Eq.~(\ref{eq:GRSALI}). 

 \section{Numerical examples} \label{sec:NumEx}

 In order to check the ability of the adjusted indicators (discussed in
 Sec.~\ref{sec:Indices}) to discern chaos from order, it is better to start with
 cases where chaos has already been found. In these cases the indicators just have 
 to verify the previous findings. Thus, the study starts with two cases of the MSM
 spacetime background, which were investigated in \cite{Han08b}.
 
\begin{figure*}[htp]
 \centerline{\includegraphics[width=0.4 \textwidth] {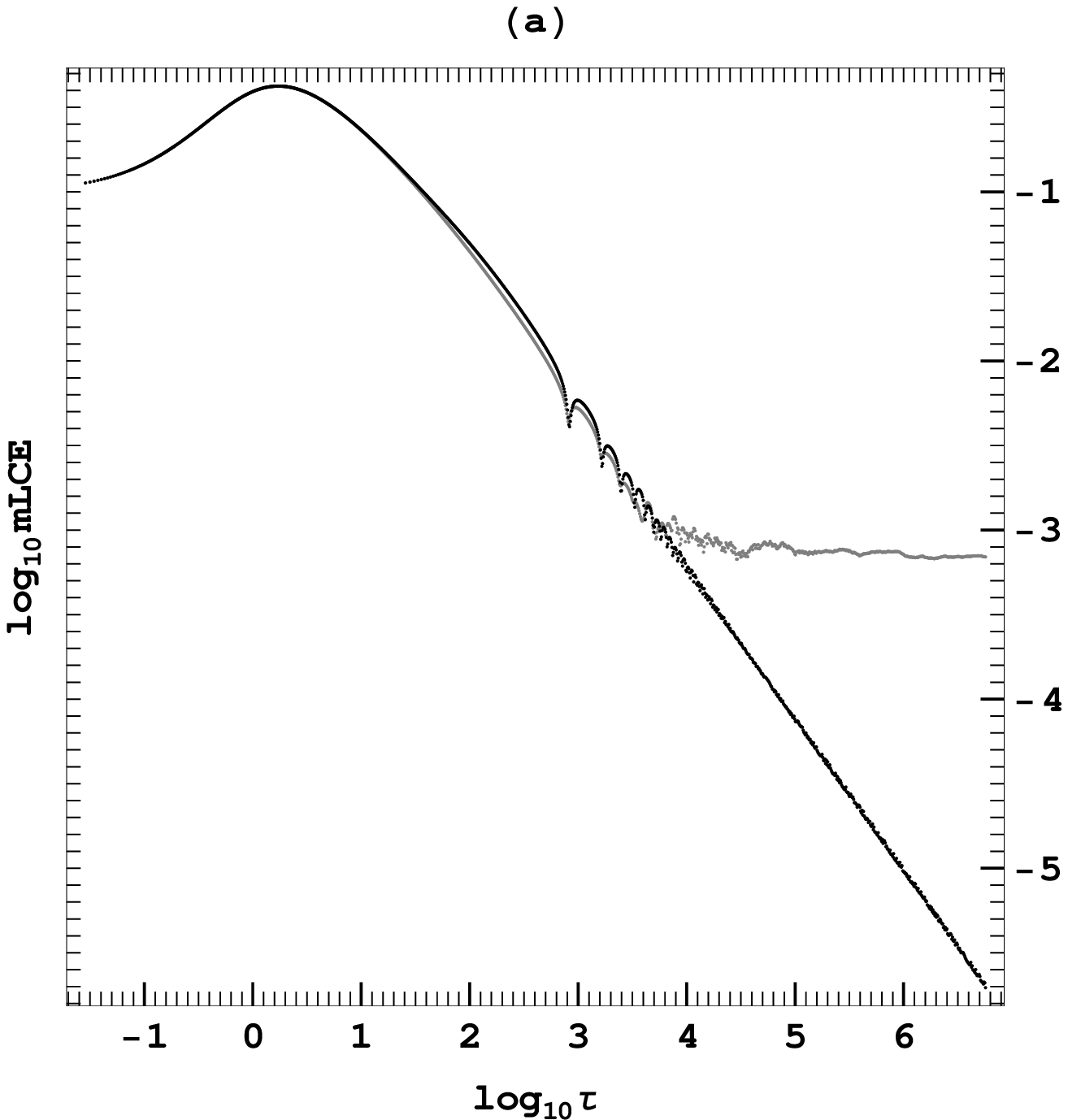}
 \includegraphics[width=0.4 \textwidth] {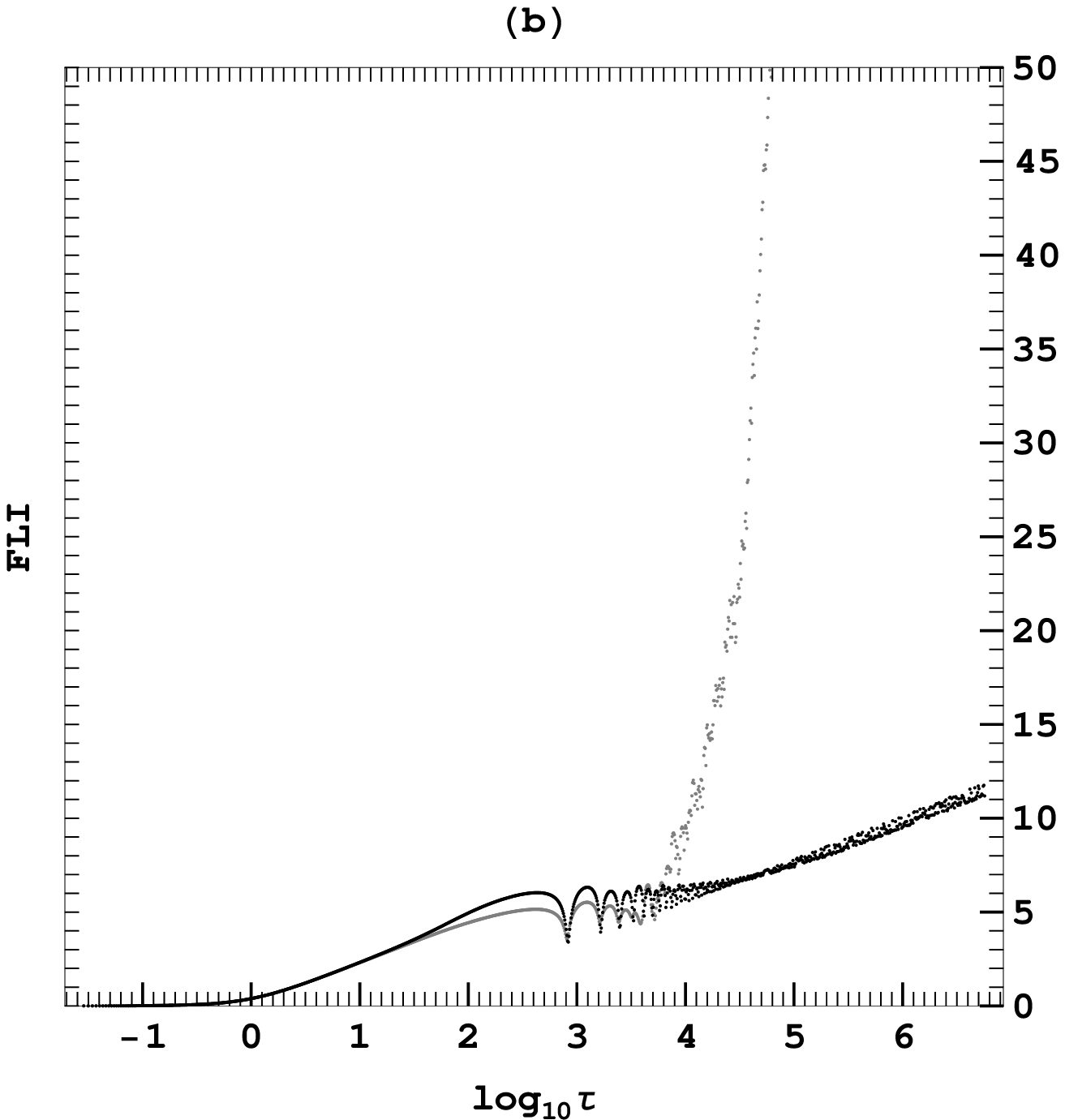}}
 \centerline{\includegraphics[width=0.4 \textwidth] {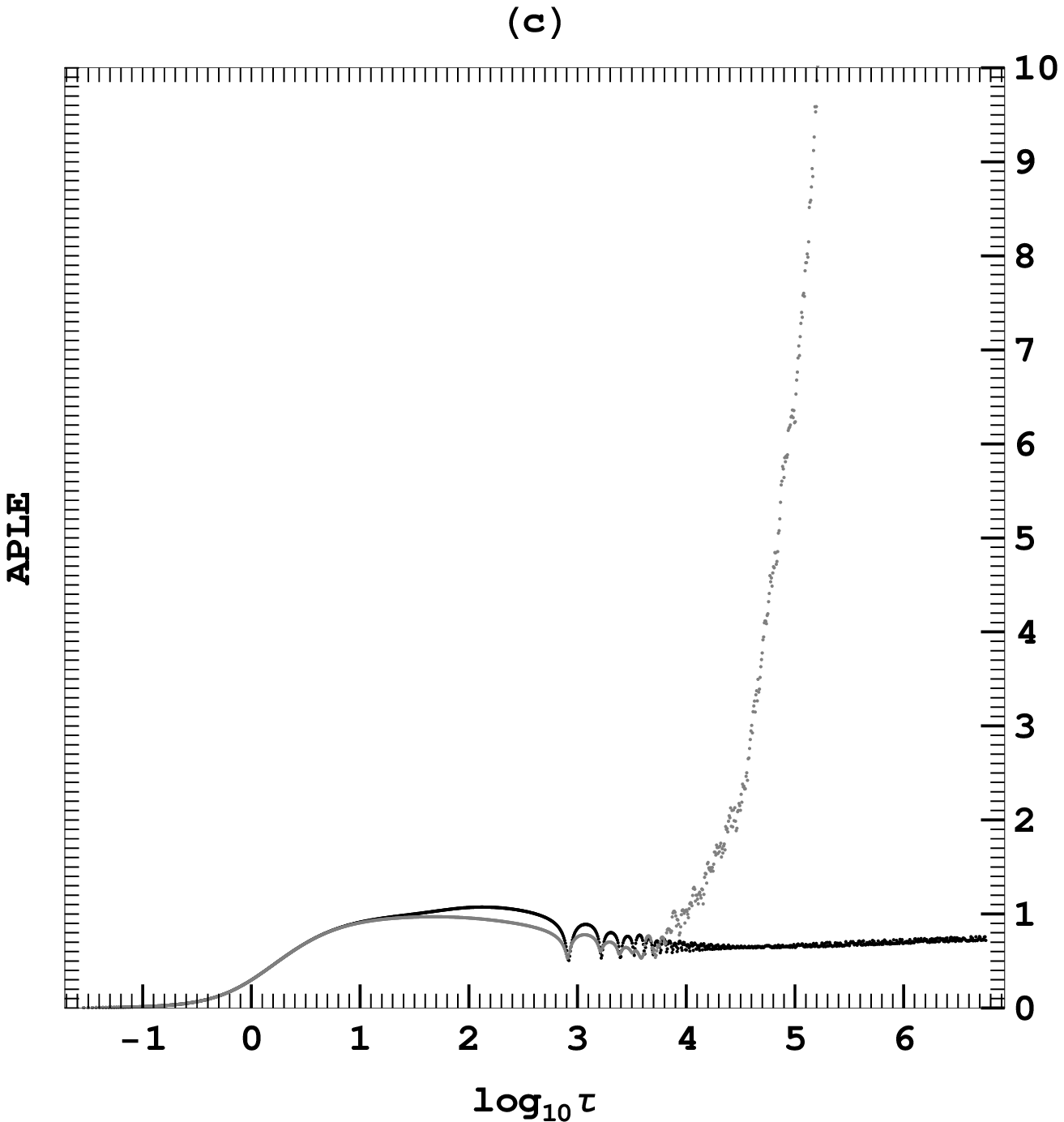}
 \includegraphics[width=0.4 \textwidth] {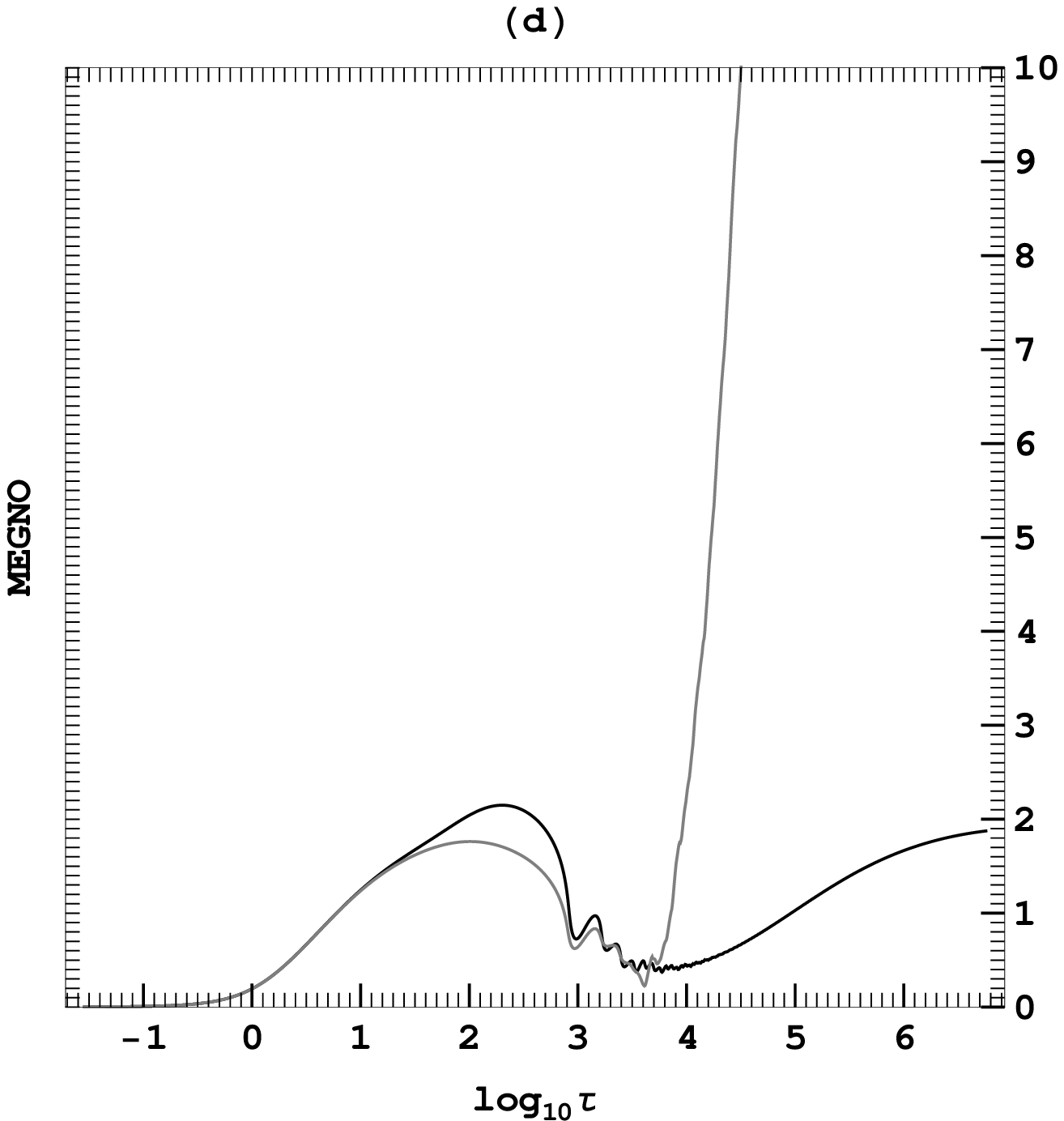}}
 \centerline{\includegraphics[width=0.4 \textwidth] {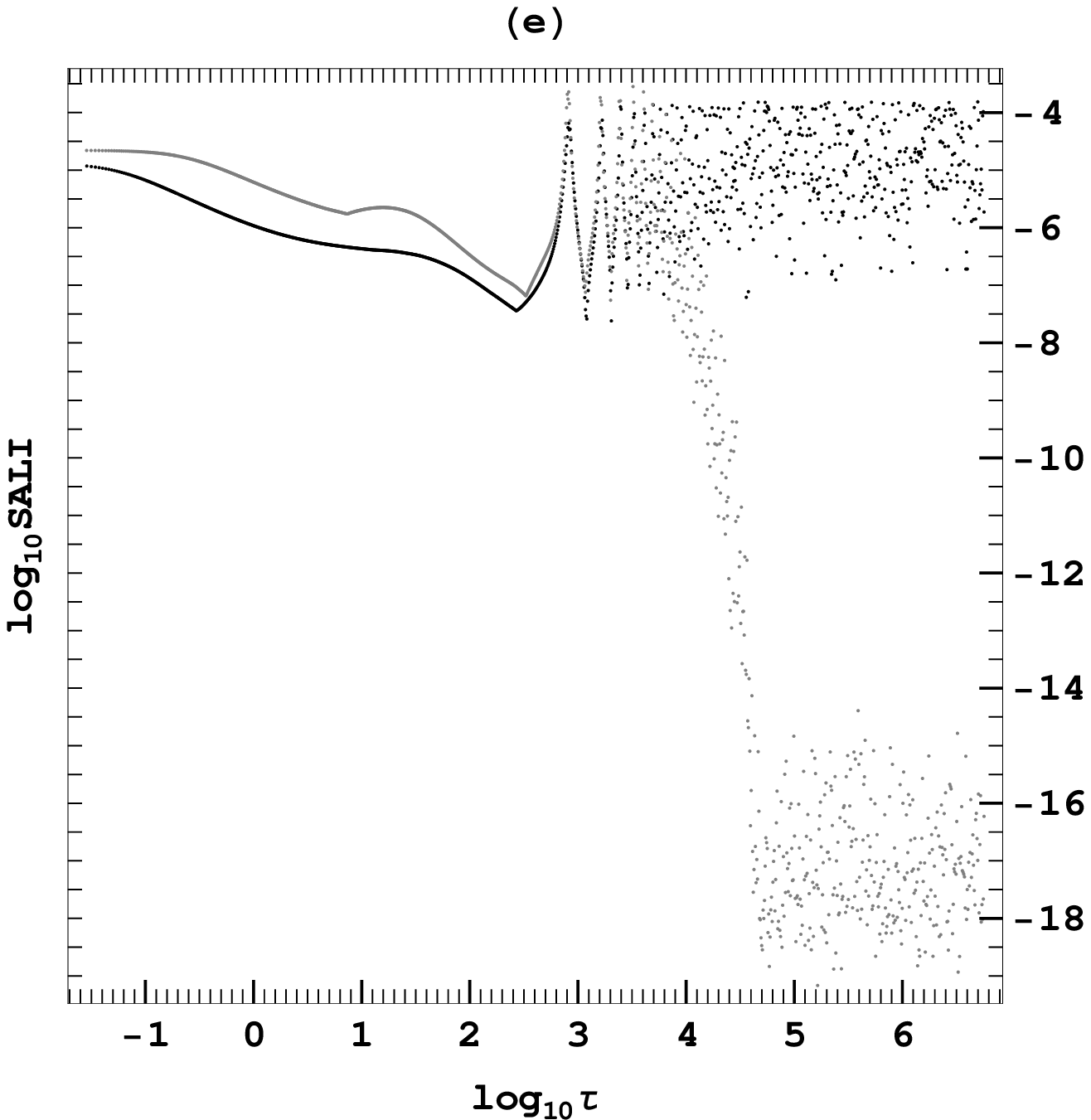}
 \includegraphics[width=0.4 \textwidth] {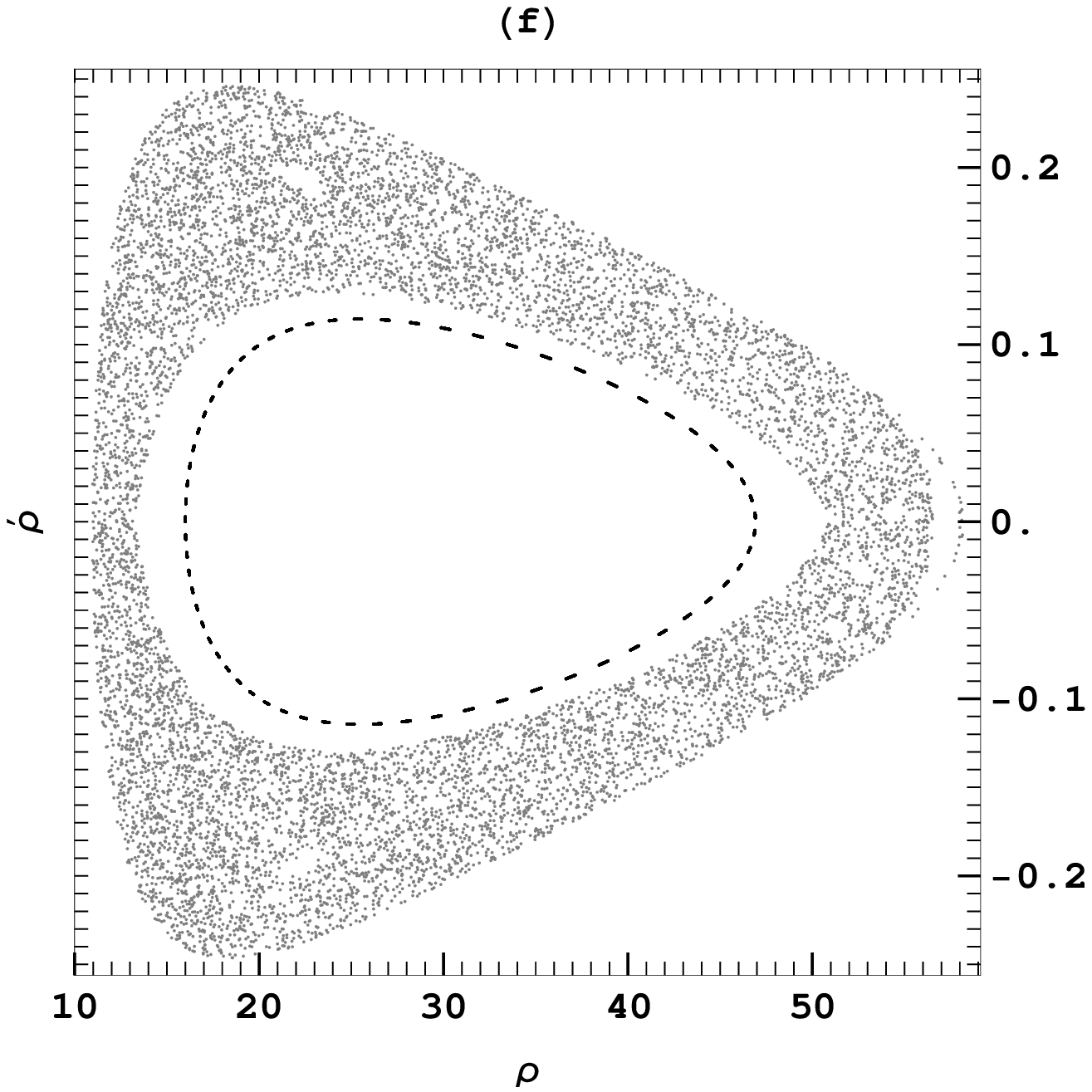}}
 \caption{ The behavior of chaotic indicators for two geodesic orbits, one regular
 (black) and one chaotic (gray), evolving in a MSM spacetime with $m=2.904$,
 $a=1.549$, $q=0$, $\mu=0$ and $b=6$. The constants of motion are $E=0.96$ and
 $L_z=2.75~m$. Panel (a) shows the evolution of $\textrm{mLCE}$ as function of the
 proper time $\tau$ in logarithmic  scale, panel (b) shows the FLI as function of
 $\log_{10}\tau$, panel (c) shows the APLE, panel (d) the MEGNO, panel (e) the
 $\log_{10}\textrm{SALI}$ as function of $\log_{10}\tau$, and panel (f) shows the
 Poincar\'{e} section on the equatorial plane $z=0$.
}
\label{figEx20}
\end{figure*}

 The first case comes from Fig.~3 of \cite{Han08}, where the mass is $m=2.904$, the
 spin is $a=1.549$, the charge is $q=0$, and the two real parameters are $\mu=0$
 and $b=6$. In fact, for all the MSM spacetimes in this work the charge $q$ and the
 parameter $\mu$ were set to zero like in previous studies \cite{Dubeibe07,Han08b}.
 The constants of motion in this example are $E=0.96$ and $L_z=2.75~m$. In
 Fig.~\ref{figEx20}, behaviors of different chaos detection techniques are shown for
 a regular orbit (black) and a chaotic orbit (gray). The initial radial distance
 for the former orbit is $\rho=16$, and for the latter $\rho=11$, while for both
 of them $\dot{\rho}=z=0$, and the $\dot{z}$ is derived from Eq.~(\ref{eq:Lagr})
 with positive sign. The initial deviation vector used for
 Figs.~\ref{figEx20}(a)-(d) has $\xi^x=10^{-4}$, $\dot{\xi}^x=10^{-3}$, and
 $\dot{\xi}^y$ was calculated from the conditions (\ref{eq:InDevPre}), while the
 other components of the deviation vector and its derivative were set to zero. For
 the SALI in Fig.~\ref{figEx20}(e) a second deviation vector has been used, which
 initially has $\zeta^{\phi}=10^{-3}$ and $\dot{\zeta}^x=10^{-1}$; $\dot{\zeta}^y$
 was evaluated from the conditions (\ref{eq:InDevPre}), while the other components
 of the deviation vector and its derivative were set to zero. Both deviation
 vectors satisfy the conditions (\ref{eq:InDevPre}). The preservation of these
 conditions and in general the numerical accuracy of the investigation is discussed
 in the Appendix~\ref{sec:NumAc}.

 The Poincar\'{e} section ($z=0$, $\dot{z}>0$) of the orbits is shown in
 Fig.~\ref{figEx20}(f) (Fig.~3 of \cite{Han08} has more details). The chaotic orbit
 evolves in a chaotic sea (gray dots); thus we expect it to be strongly chaotic,
 while the regular orbit belongs to the resonance $9/65$ and it forms a chain of 
 small islands of stability appearing like a ``dashed'' black curve.

 In Fig.~\ref{figEx20}(a), the $\log_{10}\textrm{mLCE}$ is plotted as function of
 $\log_{10}{\tau}$. In such plots the curve of regular orbits tends to zero with a
 slop $\frac{\log_{10}\textrm{mLCE}}{\log_{10}{\tau}}\propto-1$ (see the discussion
 in Sec.~\ref{subsec:mLCE}), even if the curve of chaotic orbits can follow the
 slop $-1$ for a while, when the curve reaches the value of mLCE it becomes
 horizontal. The behavior described above is what we see in Fig.~\ref{figEx20}(a).
 Namely, the black points of the regular orbit follow the slope $-1$ as mLCE tends
 to zero, and the gray points showing the evolution of the chaotic orbit follow the
 $-1$ slope for a while, but after the time $\tau\approx10^{-3}$ they change their
 inclination and become horizontal indicating the corresponding mLCE value
 ($\log_{10}\textrm{mLCE} \approx -3.15$). 

 The black points of the regular orbit in Fig.~\ref{figEx20}(b) show the
 anticipated linear growth of the corresponding deviation vector; i.e., we can see
 that $\textrm{FLI}\propto \log_{10}\tau$. The oscillations in FLI's value come
 from the fact that the tori on which the regular orbits are evolving are not in
 general direct products of circles, but rather products of ellipses; thus, the
 deviation vector's components stress and shrink periodically (for more details
 on these oscillations see e.g., the discussion in \cite{LGVE08}). On the other
 hand, the gray points of the chaotic orbit, after a certain period that they
 behave similarly to the regular orbit, begin to diverge from the regular behavior
 with time because the exponential growth of the deviation vector dominates. Thus,
 until the time of this divergence we cannot distinguish a chaotic orbit from a
 regular one. The level a regular orbit reaches at a certain time indicates the
 threshold above which we can characterize an orbit as chaotic or regular
 (Sec.~\ref{subsec:FLI}). However, this threshold is not only time-dependent, but 
 also a little bit arbitrary because we have to include a safety margin for the
 oscillations (see Sec.~\ref{subsec:FLI} and discussion in \cite{LGVE08}).     

 Examples of indicators with a time-independent threshold are the APLE and the
 MEGNO. These indicators for regular orbits tend asymptotically to $1$, and $2$,
 respectively (black points in Figs.~\ref{figEx20}(c)-(d)). In our examples,
 (Figs.~\ref{figEx20}(c)-(d)) the indicators tend to their asymptotic values from
 below (smaller values than the threshold); however, this is not always the case and
 the asymptotic behavior may be from above (see,  e.g., \cite{LG12}). Moreover, we
 have to take into account the oscillations of the deviation vector as we did for 
 FLI. Thus, it is better to set higher values than the theoretical values to these
 thresholds, in order not to characterize regular orbits as chaotic. The actual
 thresholds' values are usually set empirically, but they are not much higher than
 the theoretical ones. Now, for chaotic orbits the values of APLE and MEGNO tend to
 infinity, which is the case for the corresponding gray points shown in
 Fig.~\ref{figEx20}(c)-(d).  

 SALI differs from the other 4 indicators not only by the fact that it doesn't 
 take advantage of the deviation vector's growth (in fact SALI kills this growth
 by normalizing the components of the deviation vectors), but also by the fact that
 it needs two deviation vectors with different initial orientations in order to
 distinguish regular from chaotic orbits. For regular orbits SALI oscillates around
 a non-zero value (black dots in Fig.~\ref{figEx20}(e)), while for chaotic orbits
 SALI initially also oscillates around a non-zero value, but afterwards it plunges
 to zero (gray dots in Fig.~\ref{figEx20}(e)). The oscillations
 ($10^{-14} \lesssim SALI \lesssim 10^{-19}$) for the chaotic orbit at large
 values of proper time in Fig.~\ref{figEx20}(e) are artificial, and they result
 from numerical round offs in the summation of Eq.~(\ref{eq:GRSALI}). Thus, we have
 to set a quite arbitrary semi-empirical threshold, as was previously done for the
 other indicators, in order to characterize an orbit as chaotic. For example in
 the case of Fig.~\ref{figEx20}(e) this could be set to $10^{-10}$.  

\begin{figure*}[htp]
 \centerline{\includegraphics[width=0.4 \textwidth] {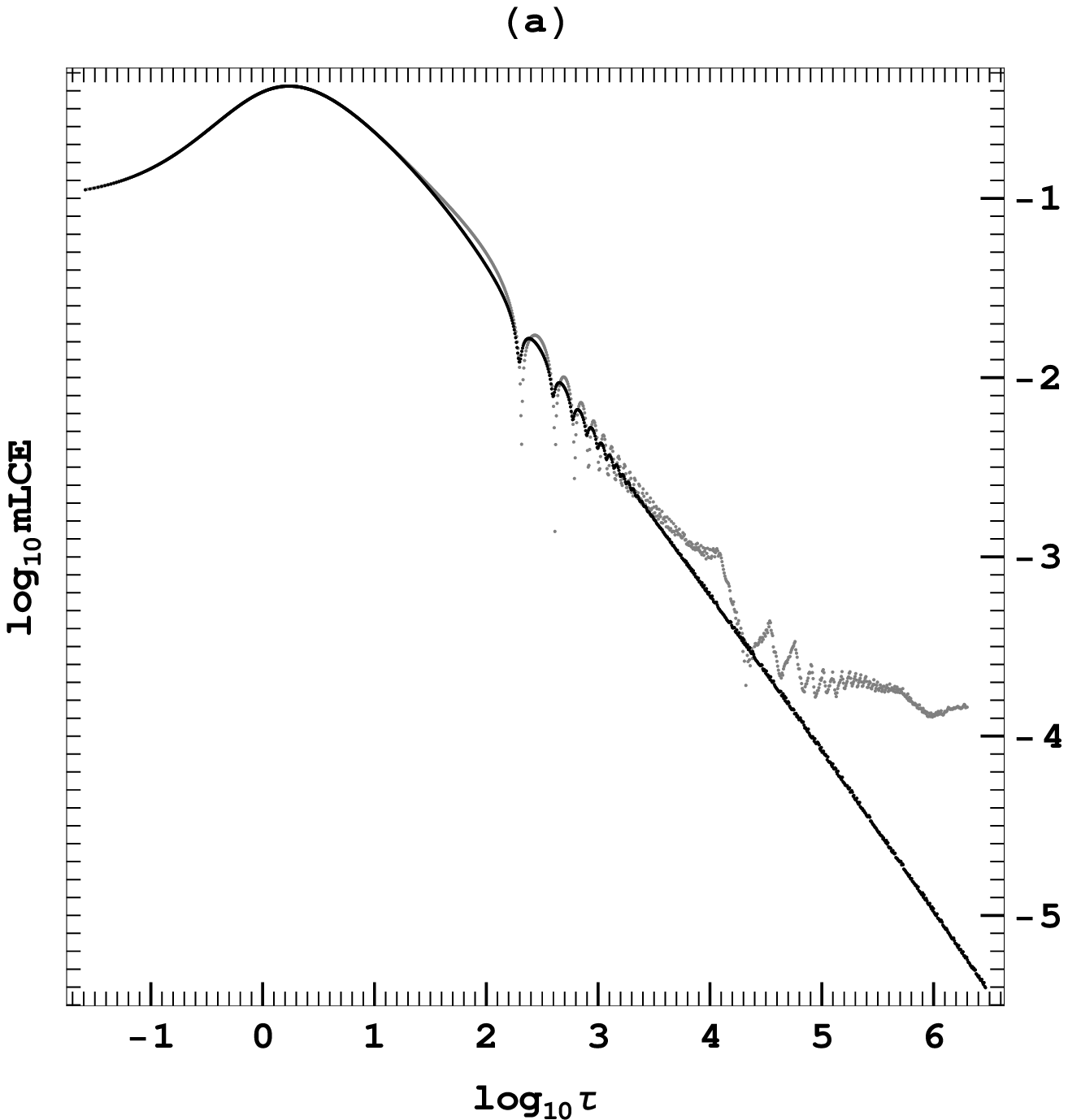}
 \includegraphics[width=0.4 \textwidth] {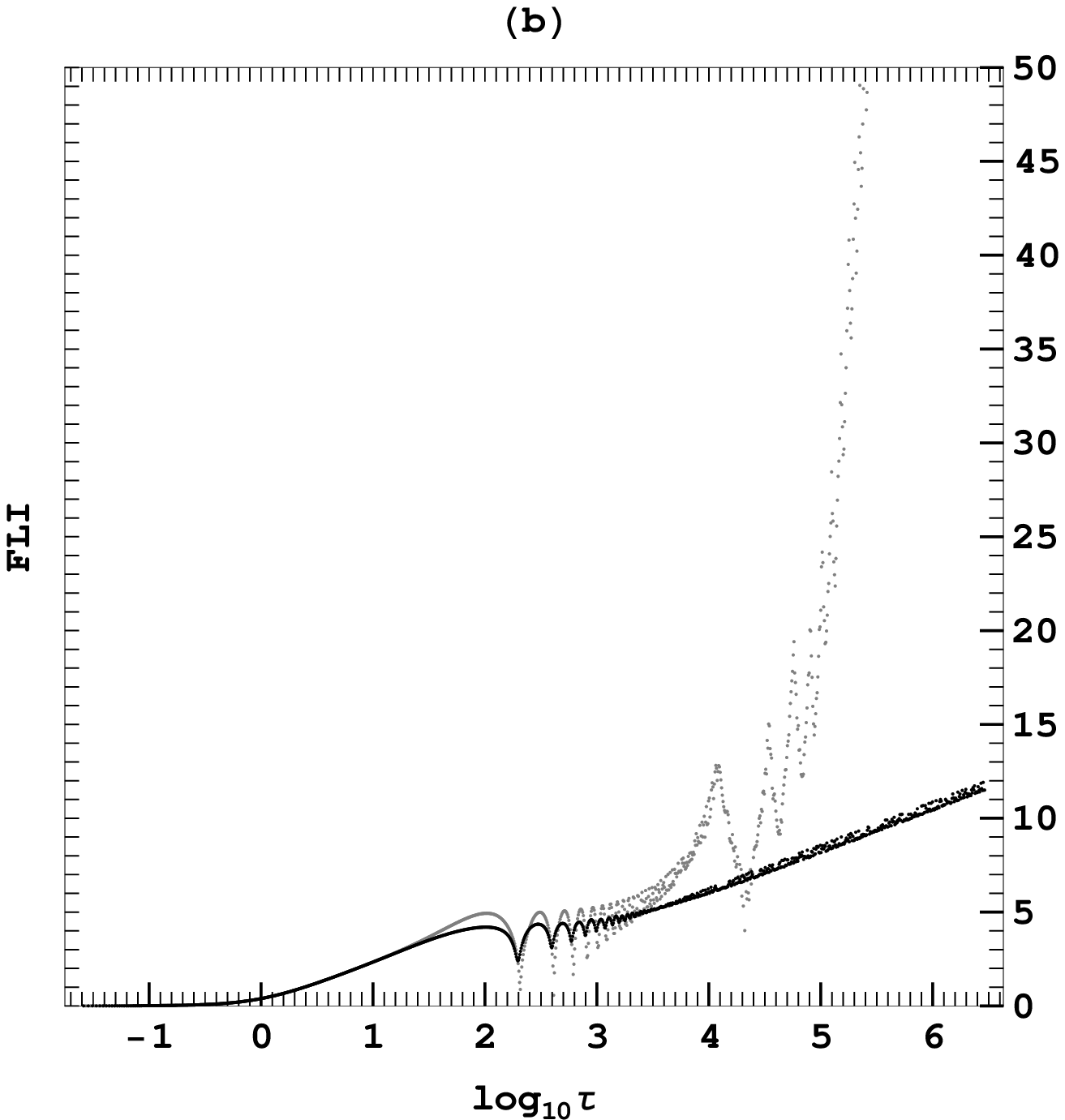}}
 \centerline{\includegraphics[width=0.4 \textwidth] {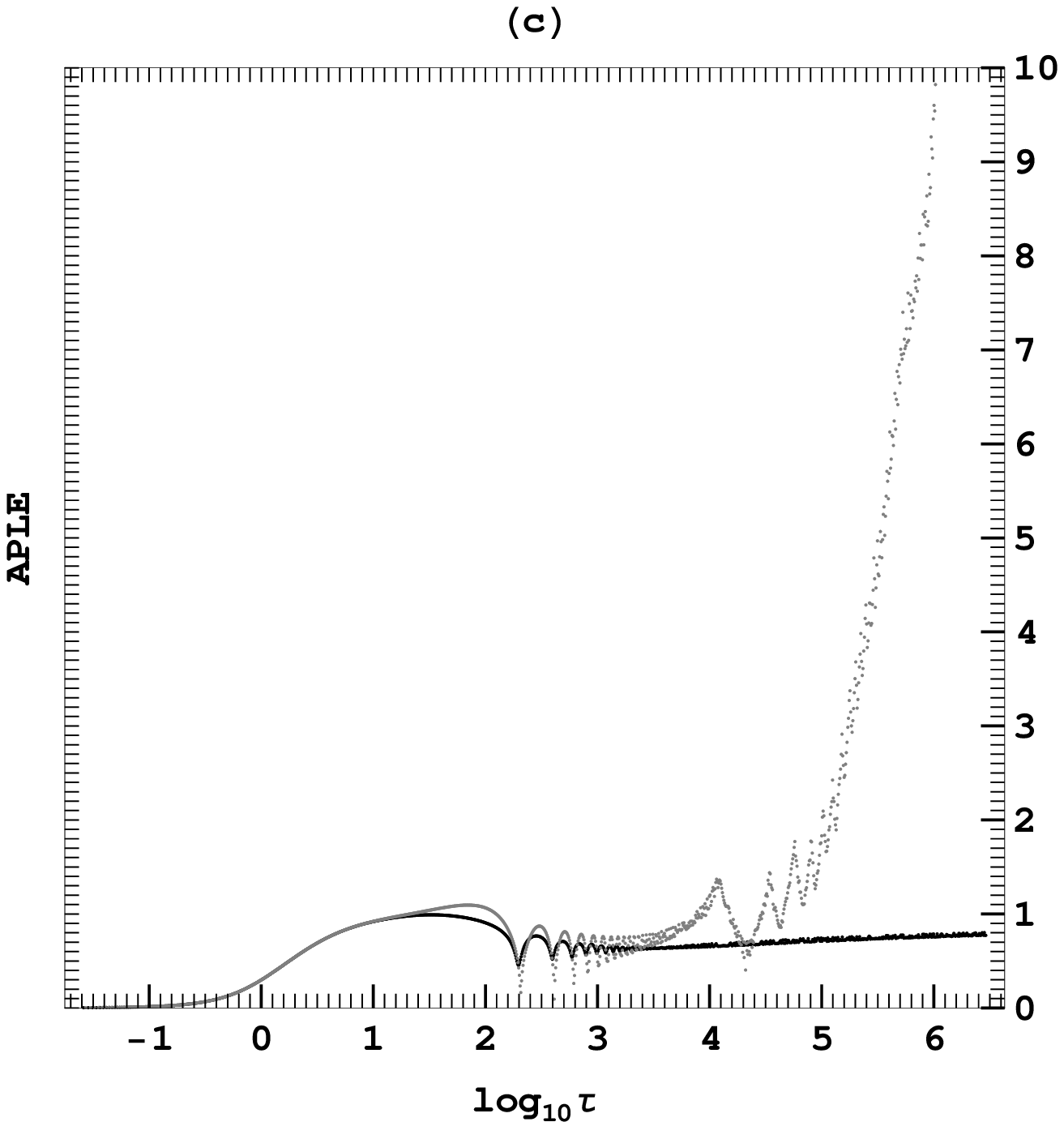}
 \includegraphics[width=0.4 \textwidth] {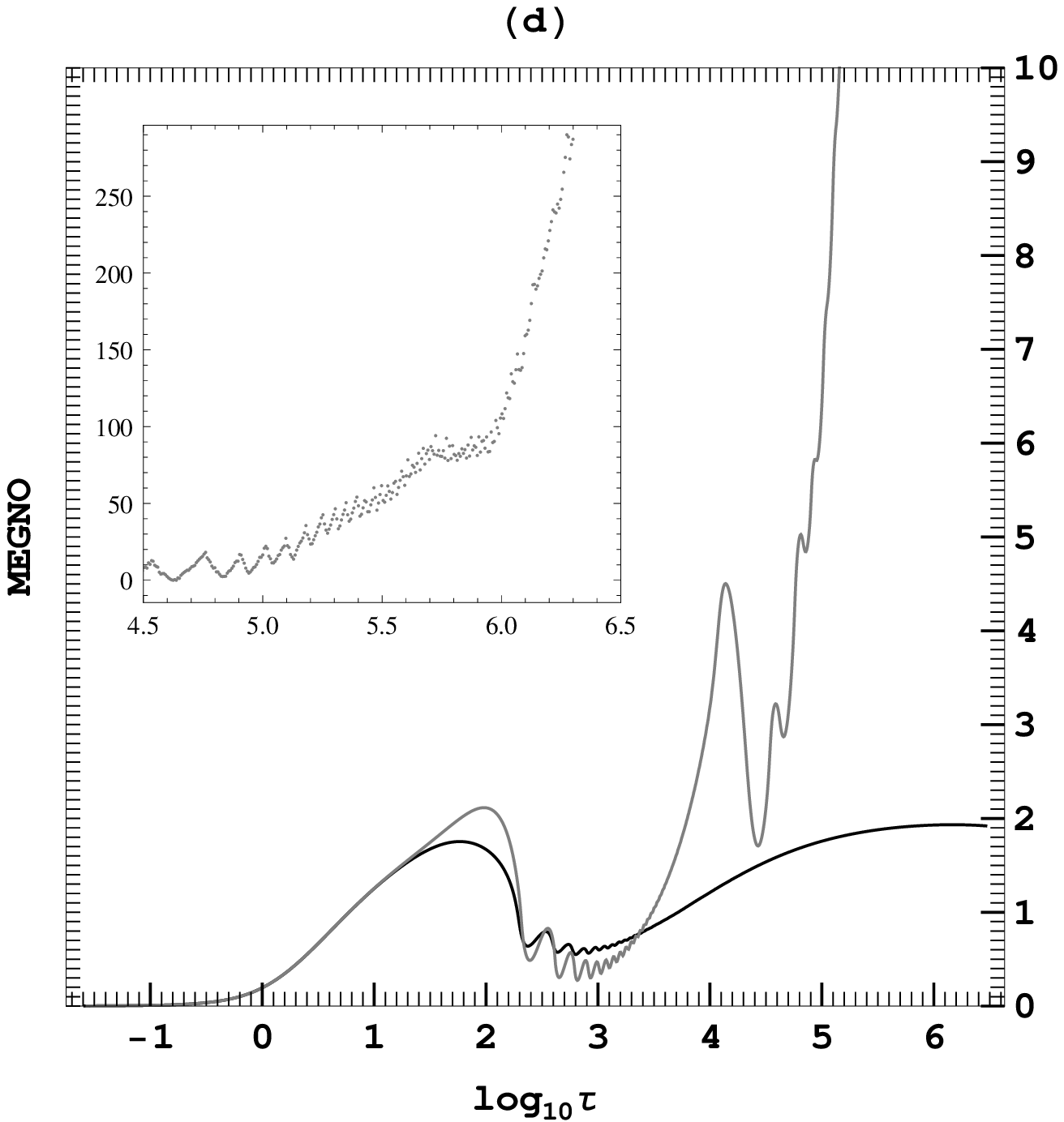}}
 \centerline{\includegraphics[width=0.4 \textwidth] {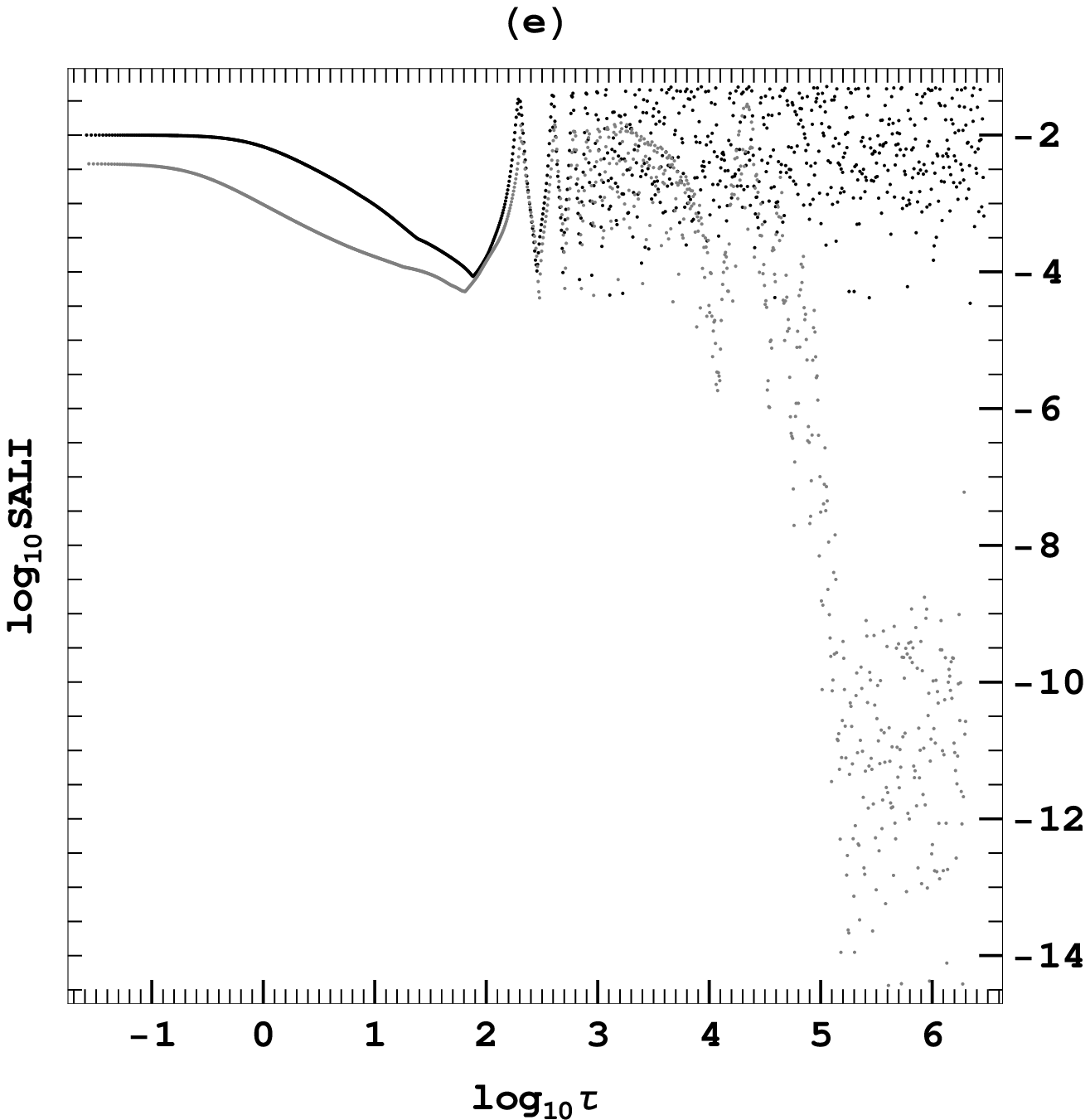}
 \includegraphics[width=0.4 \textwidth] {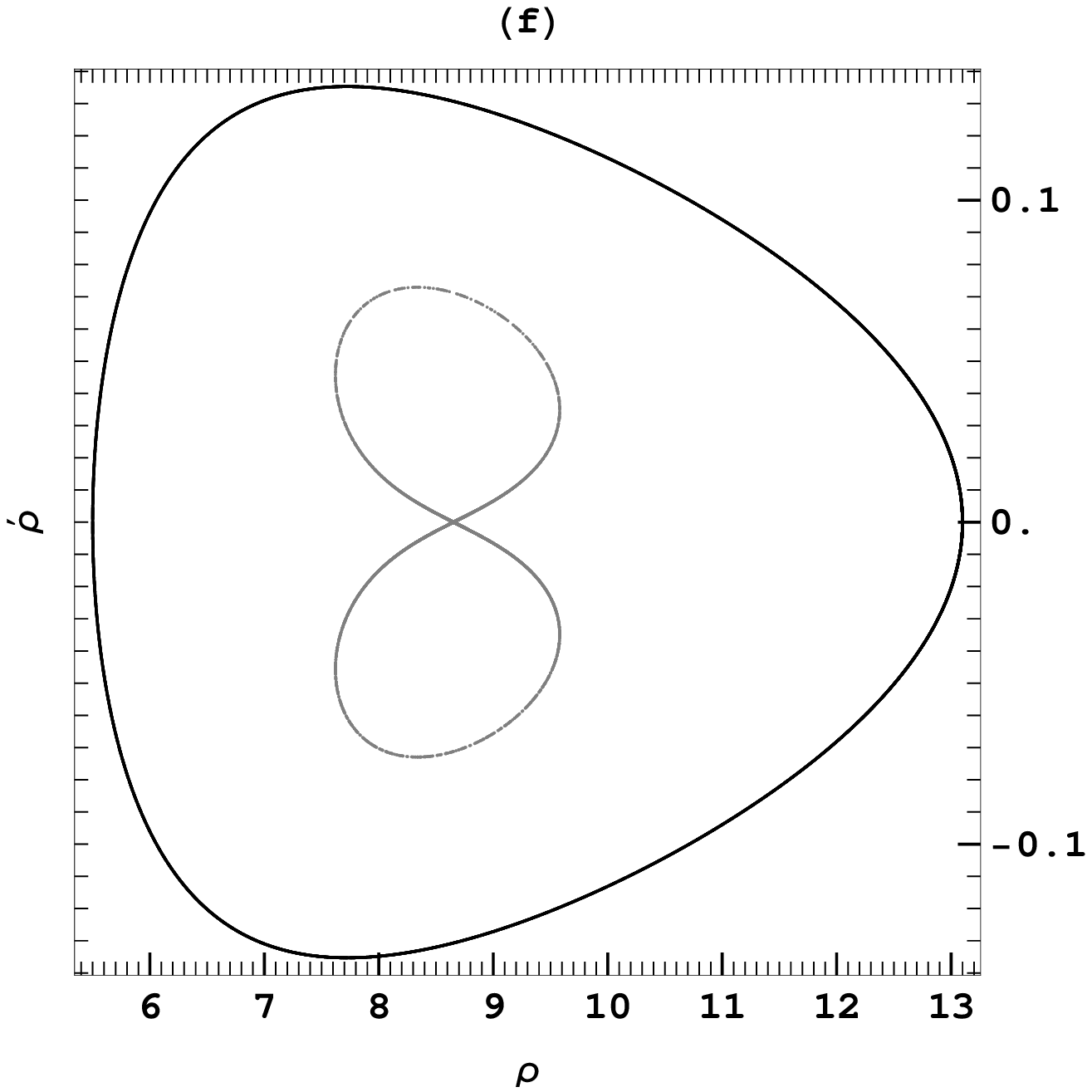}}
 \caption{The behavior of chaotic indicators for two geodesic orbits, one regular
 (black) and one chaotic (gray), evolving in a MSM spacetime with $m=1.$, $a=0.6$,
 $q=0$, $\mu=0$ and $b=3$. The constants of motion are $E=0.95$ and
 $L_z=3$. Panel (a) shows the evolution of $\textrm{mLCE}$ as function of the
 proper time $\tau$ in logarithmic  scale, panel (b) shows the FLI as function of
 $\log_{10}\tau$, panel (c) shows the APLE, panel (d) the MEGNO (the embedded 
 panel shows the non-averaged MEGNO), panel (e) the $\log_{10}\textrm{SALI}$ as
 function of $\log_{10}\tau$, and panel (f) shows the Poincar\'{e} section on the
 equatorial plane $z=0$.
}
\label{figEx10}
\end{figure*}

 The second example comes from Fig.~4 of \cite{Han08b}, where the parameters of the
 MSM spacetime are $m=1$, $a=0.6$, and $b=3$, while the test particle has $E=0.95$
 and $L_z=3$. The indicators seen in Figs.~\ref{figEx10}(a)-(e) were computed with
 the same initial deviation vectors' setup as in Fig.~\ref{figEx20}. The black 
 points correspond to the regular orbit with initial radial distance $\rho=5.5$,
 while the chaotic orbit has $\rho=8.65$. Both orbits started with $z=\dot{\rho}=0$,
 while $\dot{z}$ has been derived from Eq.~(\ref{eq:Lagr}). The Poincar\'{e}
 section for these orbits is shown in Fig.~\ref{figEx10}(f), the black curve shows
 a KAM, while the gray curve shows an orbit evolving in a chaotic layer inside the
 main island of stability.
 
 The mLCEs of chaotic orbits moving in chaotic layers like the one in
 Fig.~\ref{figEx10}(f) are usually smaller than the mLCEs of chaotic orbits moving
 in a chaotic sea (e.g., Fig.~\ref{figEx20}(f)), when the chaotic orbits belong to
 the same Poincar\'{e} section. It is rather coincidental that this holds also when
 we compare the mLCE (gray dots) in Fig.~\ref{figEx10}(a)) with that in
 Fig.~\ref{figEx20}(a), because the orbits in Fig.~\ref{figEx20} evolve in a
 different MSM spacetime than the orbits in Fig. \ref{figEx10}. In such layers
 chaotic orbits tend to stick for considerable intervals of time near a regular
 orbit, and to imitate its behavior, this phenomenon is called stickiness (see
 \cite{Contop02} for a review on the stickiness phenomenon). For instance, if a
 chaotic orbit moving in a chaotic layer seems to give the final value of mLCE
 (Fig.~\ref{figEx10}(a) until $\log_{10}\tau \approx 5.5$), then if the orbit gets
 sticky, the mLCE will start dropping following a slope similar to a regular orbit 
 (see the small drop in the mLCE value at $ 5.5 \lesssim\log_{10}\tau \lesssim 6 $
 in Fig.~\ref{figEx10}(a)). After the orbit leaves the sticky region mLCE grows
 again (Fig.~\ref{figEx10}(a)). Thus, the adjusted mLCE to curved spacetimes is
 able to detect fine structures in the phase space.

 Recall that FLI stands for fast Lyapunov indicator; thus, FLI has been designed
 to indicate the chaotic nature of an orbit quickly. For example, in
 Fig.~\ref{figEx10}(b) FLI has indicated that the orbit is chaotic at
 $\log_{10}\tau \approx 5$, while mLCE gives this indication at
 $\log_{10}\tau \approx 5.5$ (Fig.~\ref{figEx10}(a)), because we have to wait
 awhile until we are reassured that the mLCE has stopped dropping following the
 $-1$ inclination. However, this delay is not always the case; for example,
 Figs.~\ref{figEx20}(a),(b) show a case for which the detection needs approximately
 the same order of time, because the oscillations of FLI compel us to give a
 larger boundary to the limit for which we would characterize an orbit as chaotic
 (see previous discussions).

 APLE, and MEGNO are as quick as FLI in detecting the chaoticity of an orbit (e.g.,
 Figs.~\ref{figEx10}(b)-(d) and Figs.~\ref{figEx20}(b)-(d)), and they show the same
 sensitivity in detecting the stickiness interval. In particular, in 
 Fig.~\ref{figEx10}(c) only a small break in the rate at which APLE tends to
 infinity can be seem for $ 5.5 \lesssim\log_{10}\tau \lesssim 6 $. FLI can detect
 this stickiness interval in the same way, but the change in this inclination is
 nearly visible like for APLE in Fig.~\ref{figEx10}. However, the MEGNO without the
 averaging (Eq.~(\ref{eq:DisMEGNO})) produces an observable plateau (embedded panel
 in Fig.~\ref{figEx10}(d)), during the time the orbit is sticky. The averaging is
 the reason why this plateau disappears in the averaged MEGNO and only a break in
 the rate gives away the stickiness. Thus, like in the case of mLCE, the
 reformulated chaotic indicators discussed in this paragraph, and in particular the
 reformulated MEGNO, are able to detect fine structures.

 On the other hand, even if SALI is as fast as FLI, APLE and MEGNO in detecting
 chaos, there is no apparent evidence of stickiness in Fig.~\ref{figEx10}(e). It
 appears that once the deviation vectors become parallel, they do not diverge again.
  
 The two examples (Figs.~\ref{figEx20},~\ref{figEx10}) show that the readjusted
 indicators have the behavior which we would expect from their classical definition.
 In general, all the indicators have the same time response in detecting chaos, and
 this time depends on the maximum Lyapunov characteristic exponent. However, each
 of them has a special ability, which can make it ideal when a specific
 investigation of a dynamical system is required; e.g., APLE was designed to detect
 power law governed metastable behaviors. However, when the only aim is chaos
 detection, the indicator one chooses is a matter of convenience and taste. 
  
\begin{figure*}[htp]
 \centerline{\includegraphics[width=0.4 \textwidth] {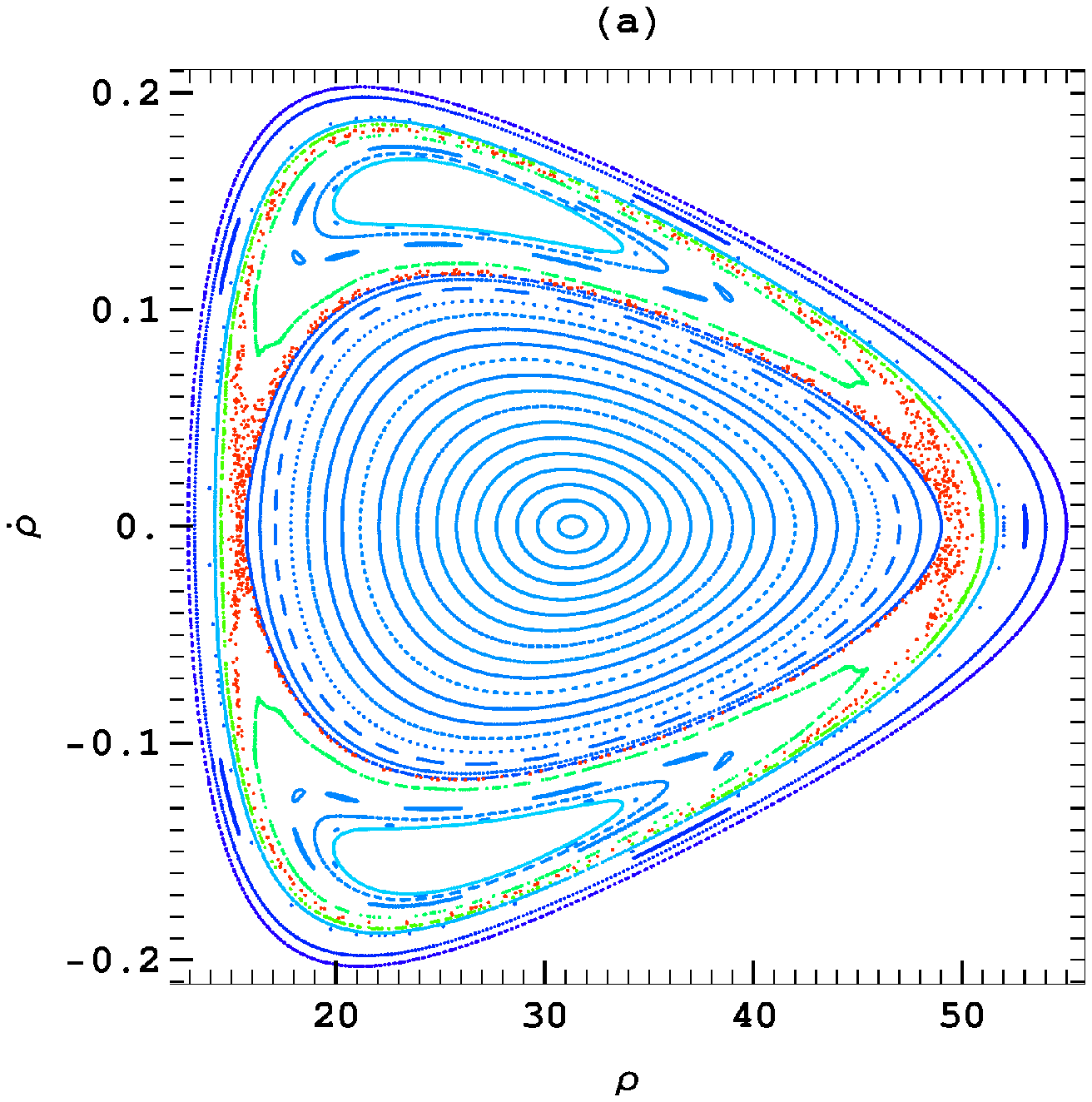}
 \includegraphics[width=0.4 \textwidth] {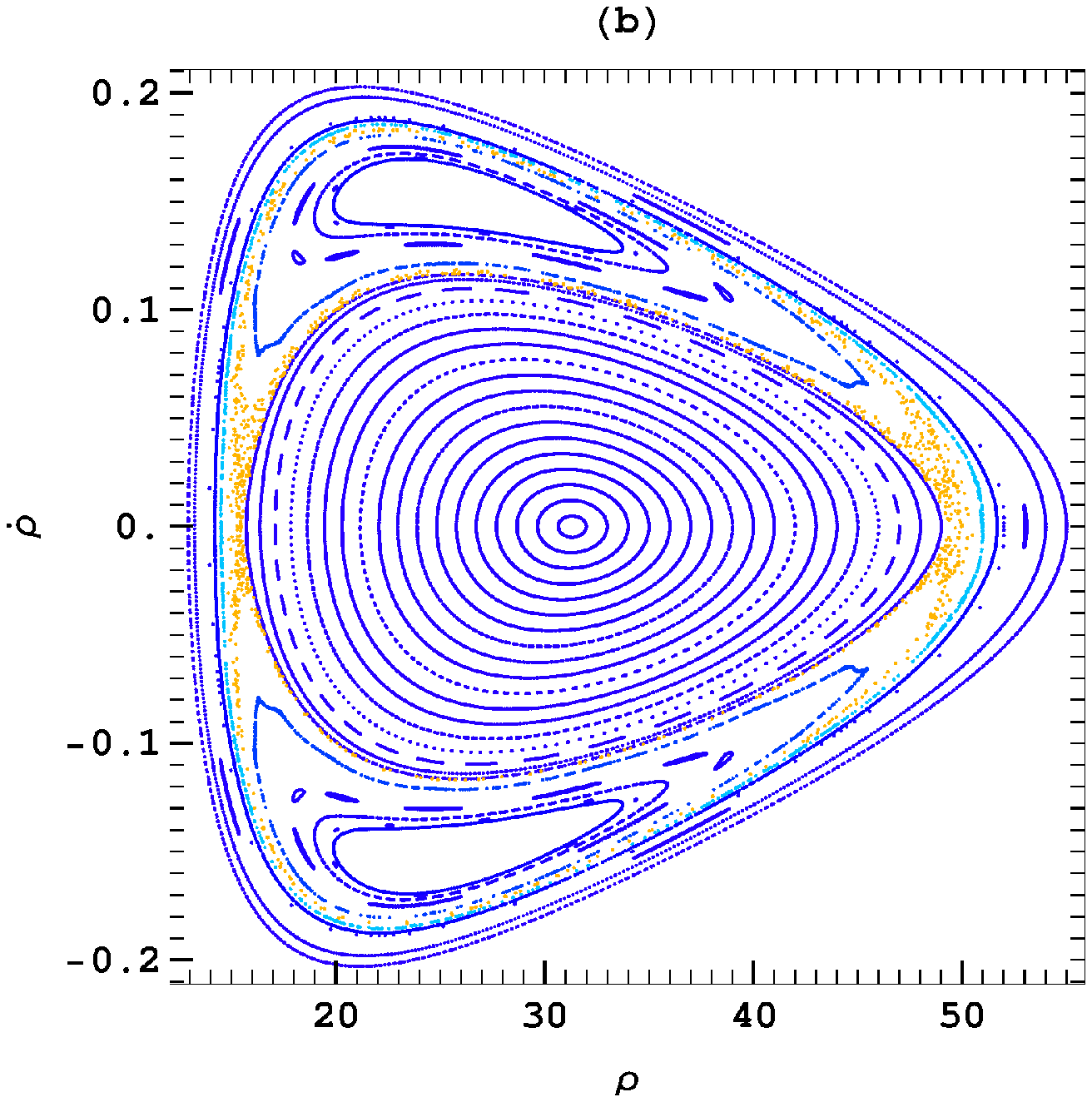}}
 \centerline{\includegraphics[width=0.4 \textwidth] {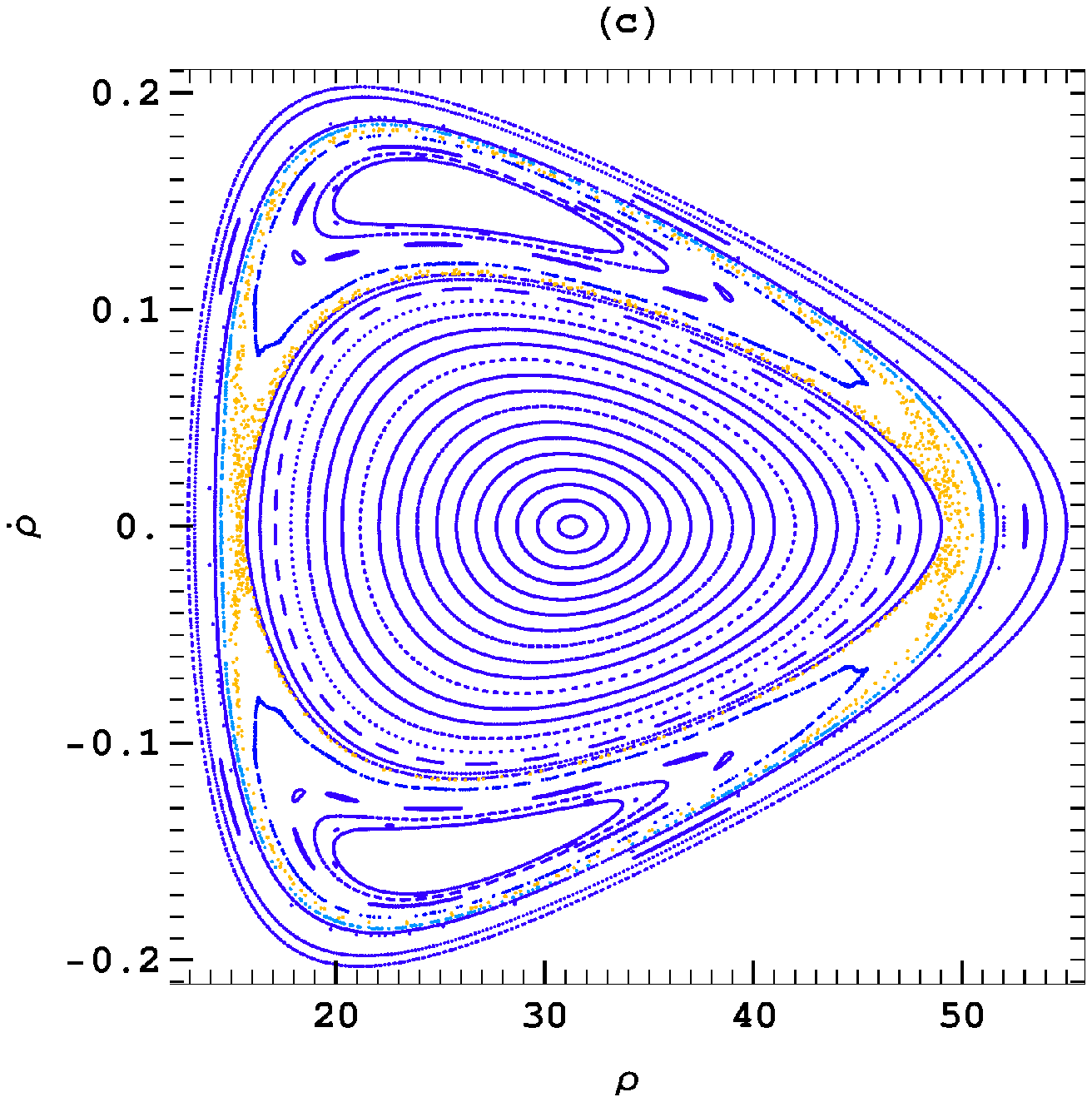}
 \includegraphics[width=0.4 \textwidth] {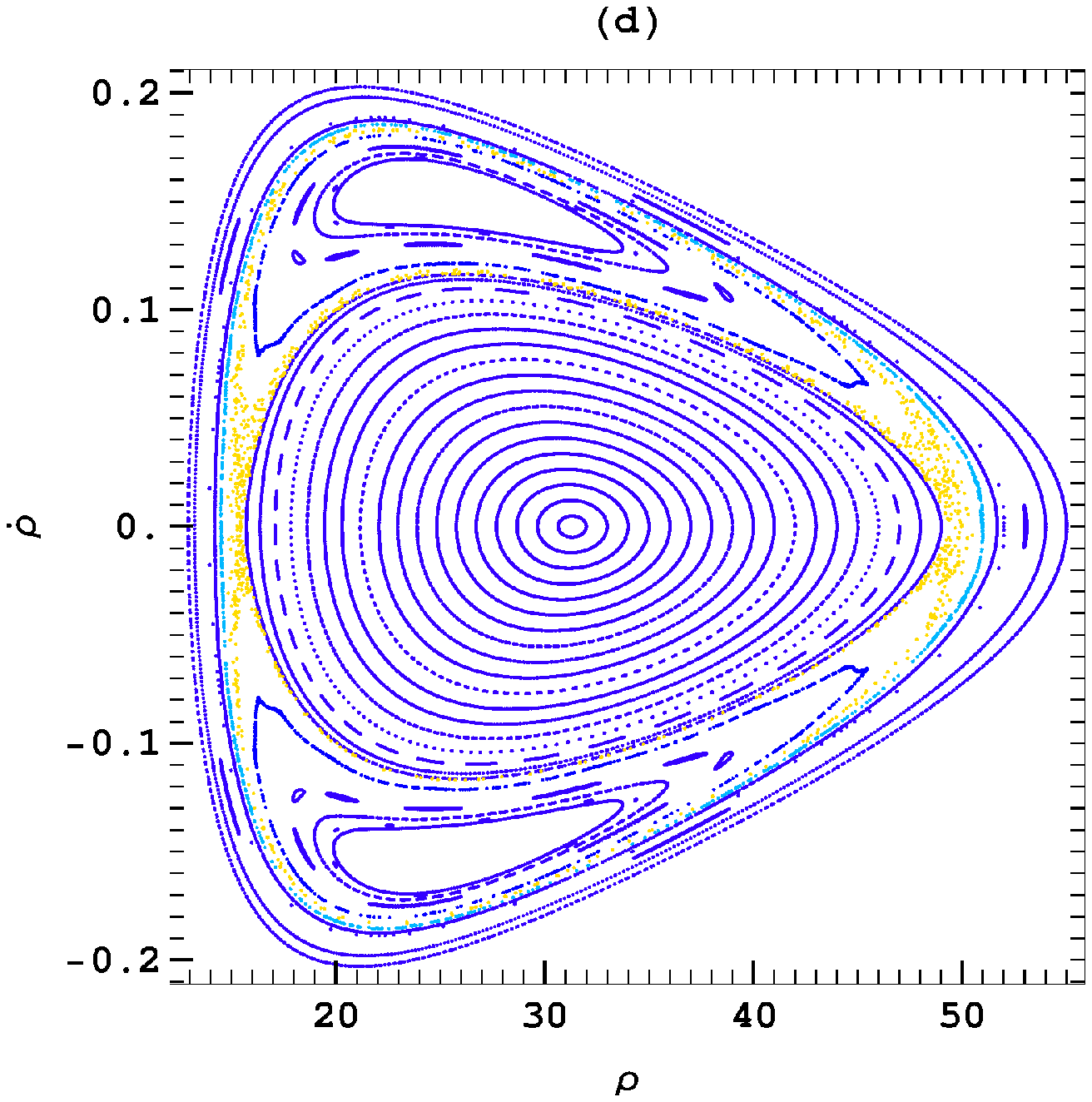}}
 \centerline{\includegraphics[width=0.4 \textwidth] {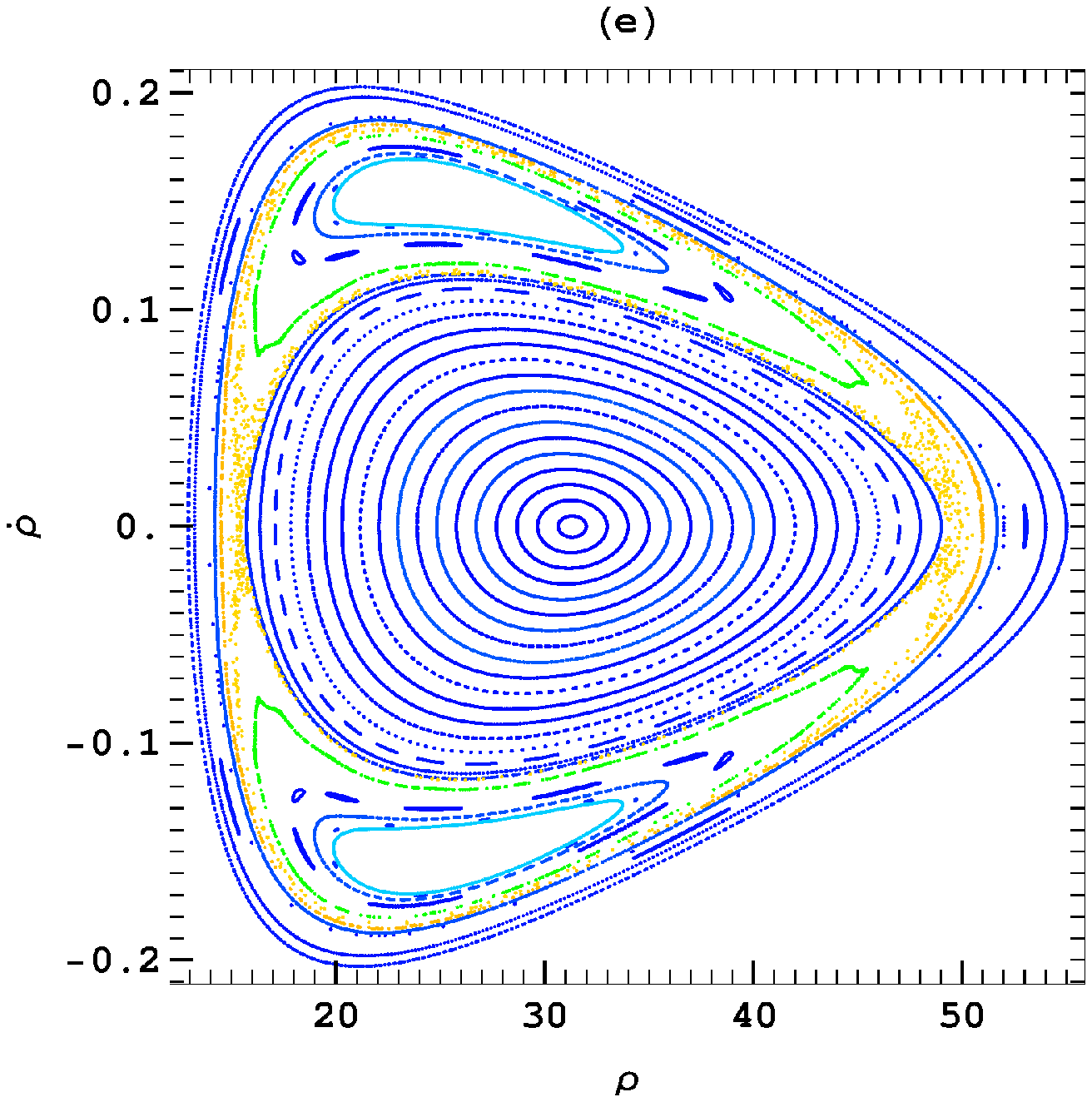}
  \hspace{0.1\textwidth}
 \includegraphics[width=0.3 \textwidth] {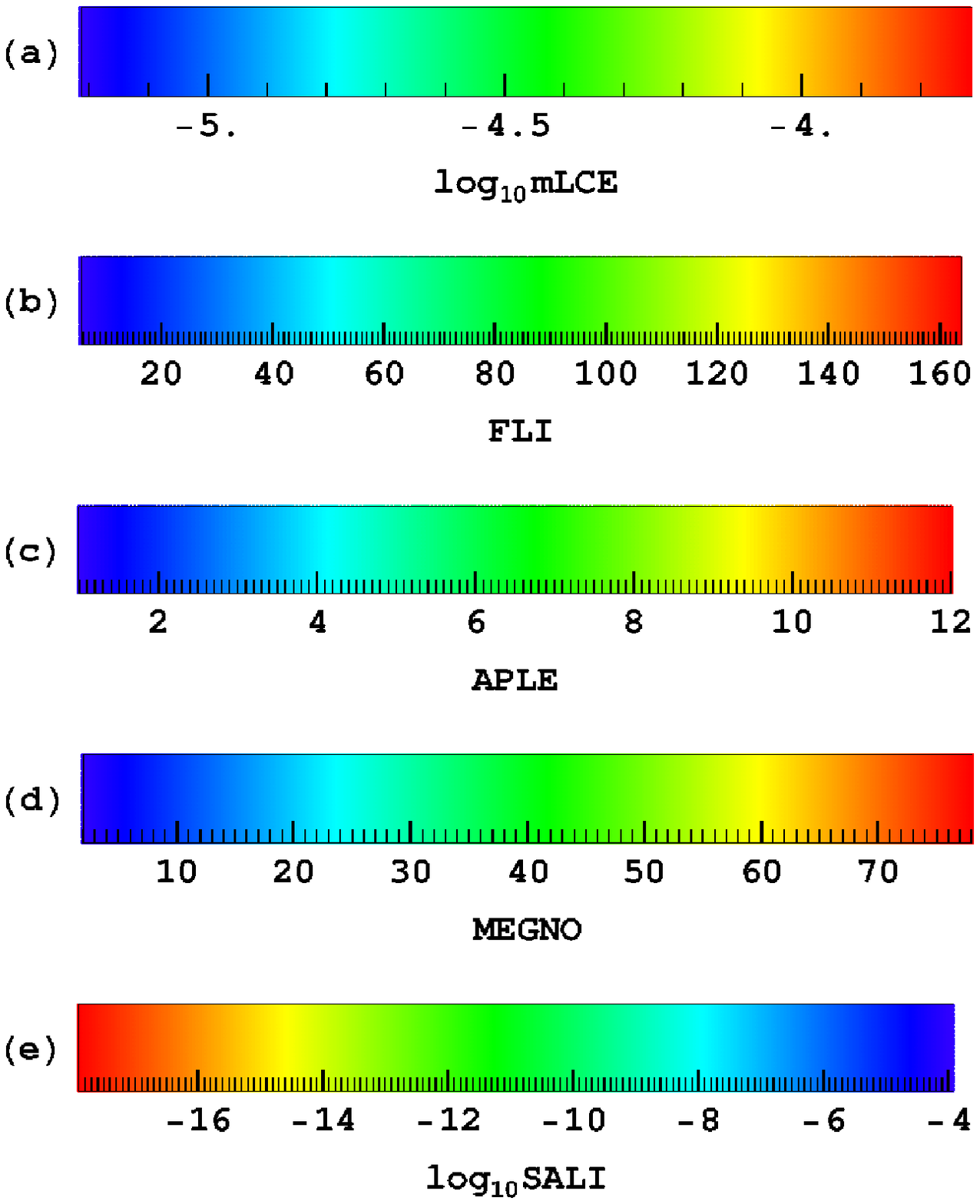}}
 \caption{The values of the chaotic indicators on the Poincar\'{e} section lying on 
 the equatorial plane $z=0$. The orbits evolve in a MSM spacetime with $m=2.904$,
 $a=1.549$, $q=0$, $\mu=0$ and $b=6$. The constants of motion are $E=0.96$ and
 $L_z=3.~m$. Panels (a)-(e) show the values of $\textrm{mLCE}$, FLI, APLE, MEGNO,
 and $\log_{10}\textrm{SALI}$ respectively in scales shown at the bottom right
 corner of the figure.
 }
\label{figEx22}
\end{figure*}

 In order to reinforce the above point, the values of the five indicators under
 discussion are plotted on a Poincar\'{e} section in Fig.~\ref{figEx22}. This
 Poincar\'{e} section lies on the equator $z=0$ and has $\dot{z}>0$. The scales of
 the indicators' values are given in the bottom right corner of Fig.~\ref{figEx22}.
 The orbits evolve in a MSM spacetime with $m=2.904$, $a=1.549$, $b=6$, and the
 constants of the motion are $E=0.96$, $L_z=3.~m$. For all orbits the initial
 deviation vectors are the same as those in Figs.~\ref{figEx20},~\ref{figEx10}, and
 all five indicators have been evaluated for $1000$ sections. All five show clearly
 which regions are dominated by chaotic orbits and which by regular motion. 

 \section{Conclusions} \label{sec:Conc} 
 
 The chaotic indicators mLCE, FLI, GALI, MEGNO, and APLE had been defined in the
 framework of classical mechanics. In order to make these indicators appropriate
 for studying geodesic motion in curved spacetimes, they have to be reformulated in
 a way that will make them invariant under coordinate transformations. The authors
 of \cite{Sota96} provided a guideline to do this when they reformulated mLCE. By
 following this guideline the other four chaotic indicators were reformulated
 accordingly in Sec.~\ref{sec:Indices}. All the five reformulated indicators were
 tested in Sec. \ref{sec:NumEx} for their efficiency in discerning regular from
 chaotic motion in the MSM spacetime background. It was shown that these five
 indicators have inherited the anticipated behavior from their classical
 counterparts, they are reliable, and, in general, equally fast in detecting
 chaotic motion.

\begin{acknowledgments}
 G.~Lukes-Gerakopoulos was supported by the DFG grant SFB/Transregio 7. I would
 like to thank Bernd Br\"{u}gmann, Tim Dietrich, and Jonathan Seyrich for their
 suggestions.
\end{acknowledgments}

\appendix

\section{Numerical accuracy} \label{sec:NumAc}

 For integrating the geodesics Eq.~(\ref{eq:geod}) and their deviations
 Eq.~(\ref{eq:GeoDev}) the IGEM integration scheme \cite{Seyrich12} was implemented.
 This scheme was designed to cope with strongly chaotic geodesic motion efficiently
 and accurately. In this section we discuss IGEM's performance.

 One point not included in our previous analysis \cite{Seyrich12} is
 the renormalization of evaluated quantities during the evolution. The
 renormalization is applied in order to avoid the occurrence of very large numbers,
 which aside from causing other problems would slow down the IGEM integration
 scheme. These very large numbers appear due to the fact that the measure of the
 deviation vector grows exponentially; thus, in order to get rid of the growth of 
 the corresponding values we renormalize them. In fact the renormalization of the
 deviation vector is a common ground in dynamical studies. Hence, anytime $\Xi$
 becomes larger than $10$, the components of the deviation vector and the
 components of its velocity are multiplied by $10^{-2}$. However, there are times
 at which the measure of $\Xi$ can become very small; in such cases IGEM would
 choose its integrating step by taking into account mainly the needs of the
 geodesic orbit. In order to avoid this, anytime $\Xi$ drops below $10^{-3}$, the
 aforementioned components are divided by $10^{-2}$. Thus, by renormalization, the 
 IGEM scheme is kept accurate and fast.

\begin{figure*}[htp]
 \centerline{\includegraphics[width=0.4 \textwidth] {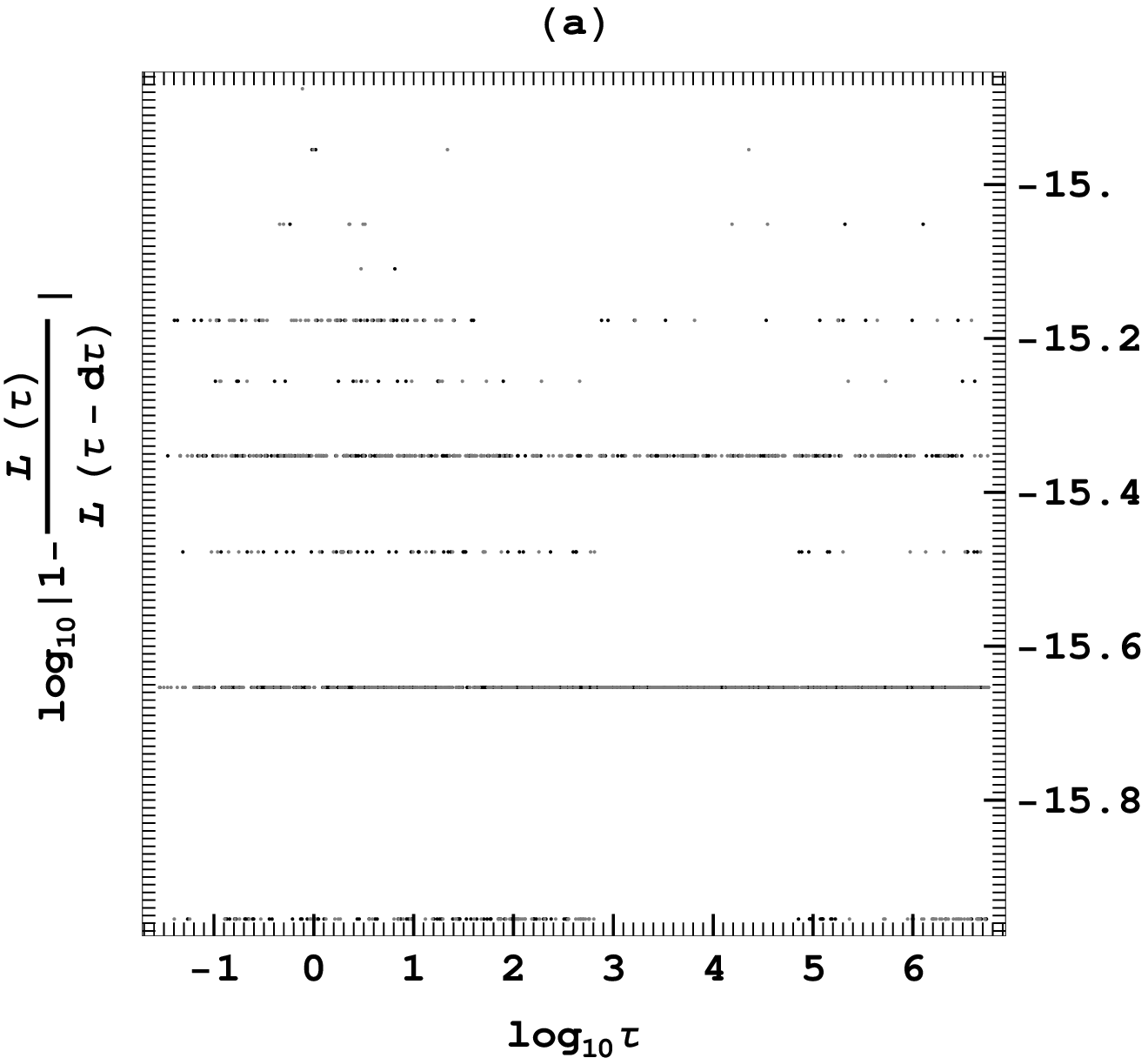}
 \includegraphics[width=0.4 \textwidth] {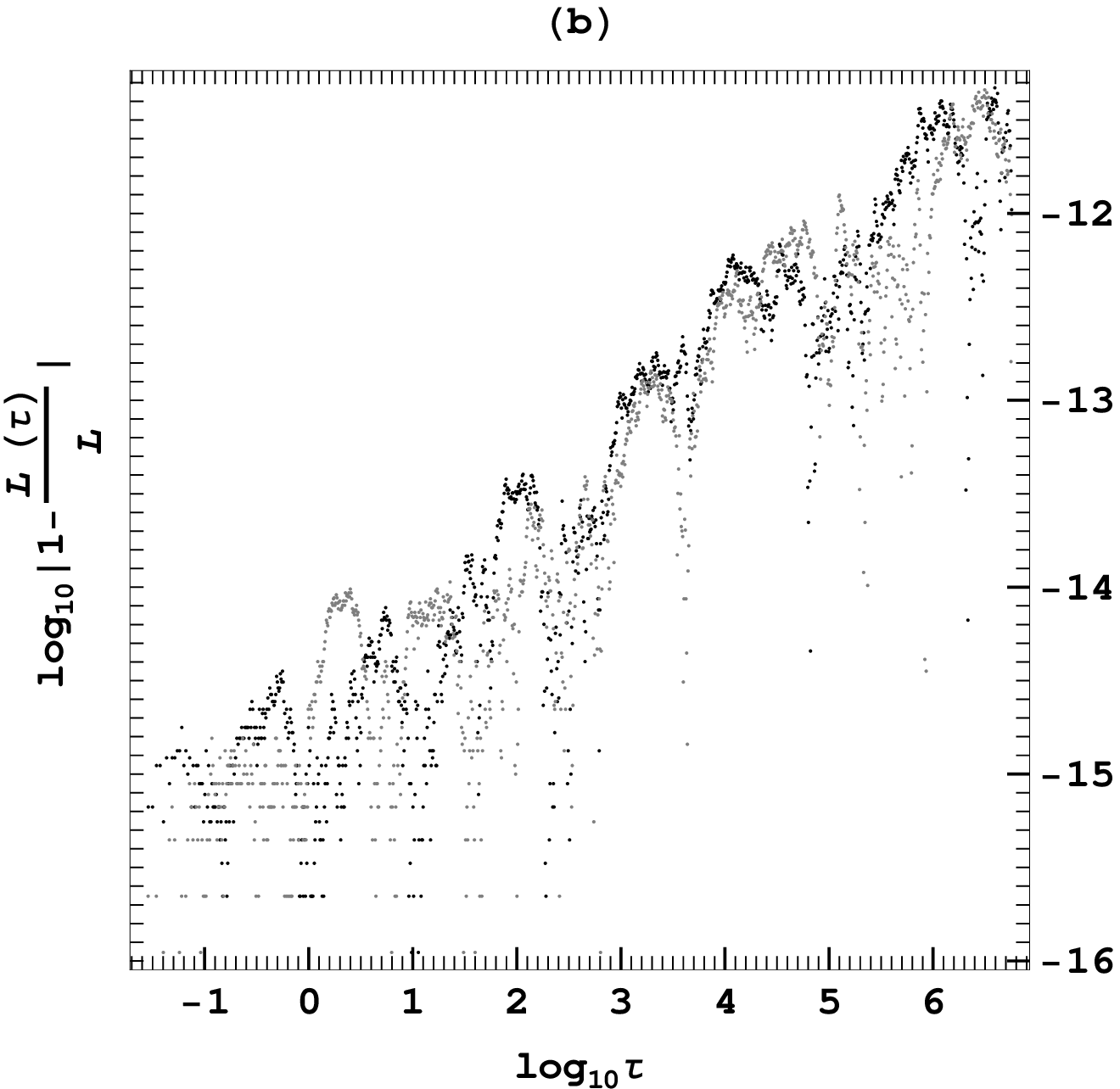}}
 \centerline{\includegraphics[width=0.4 \textwidth] {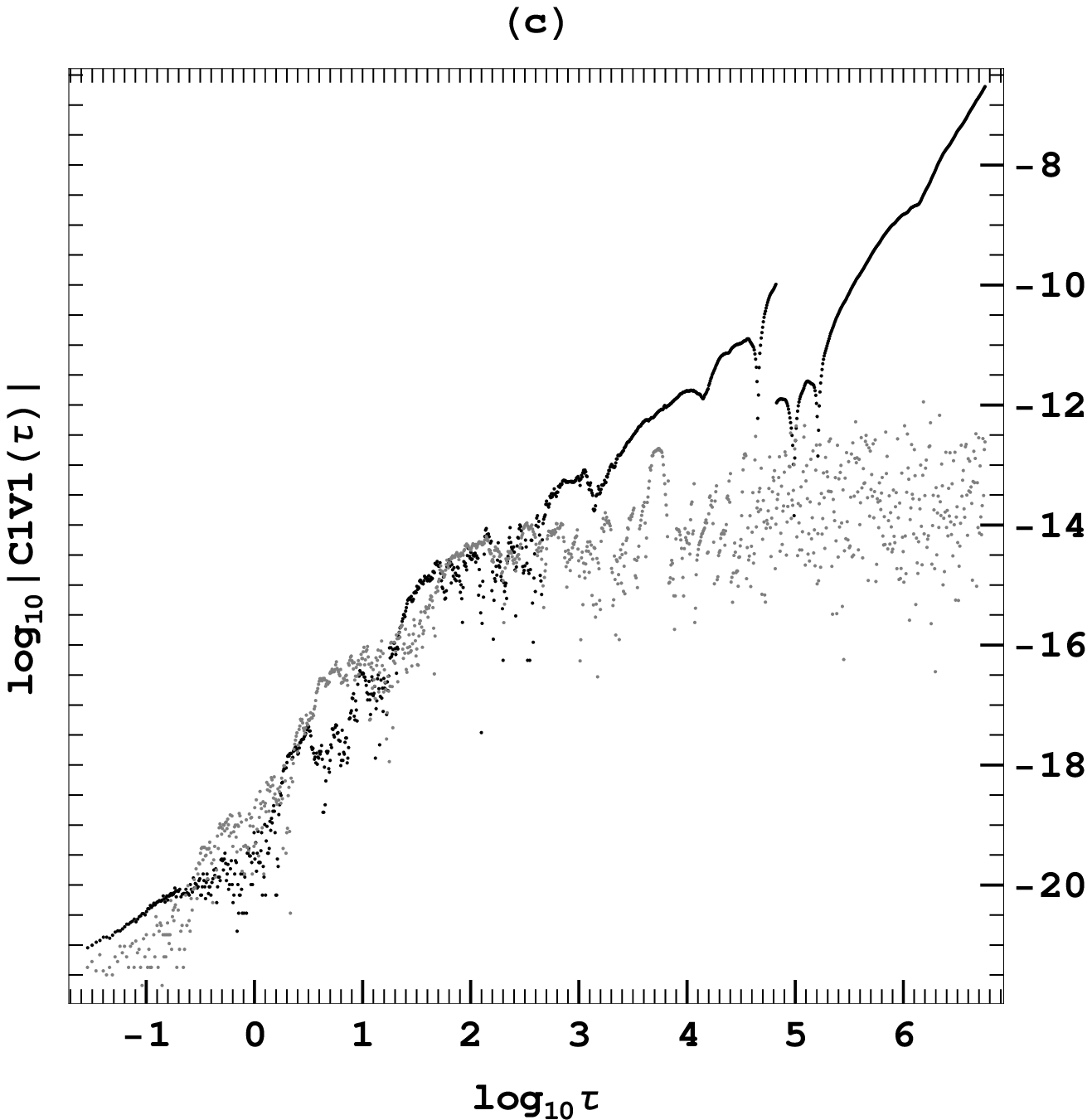}
 \includegraphics[width=0.4 \textwidth] {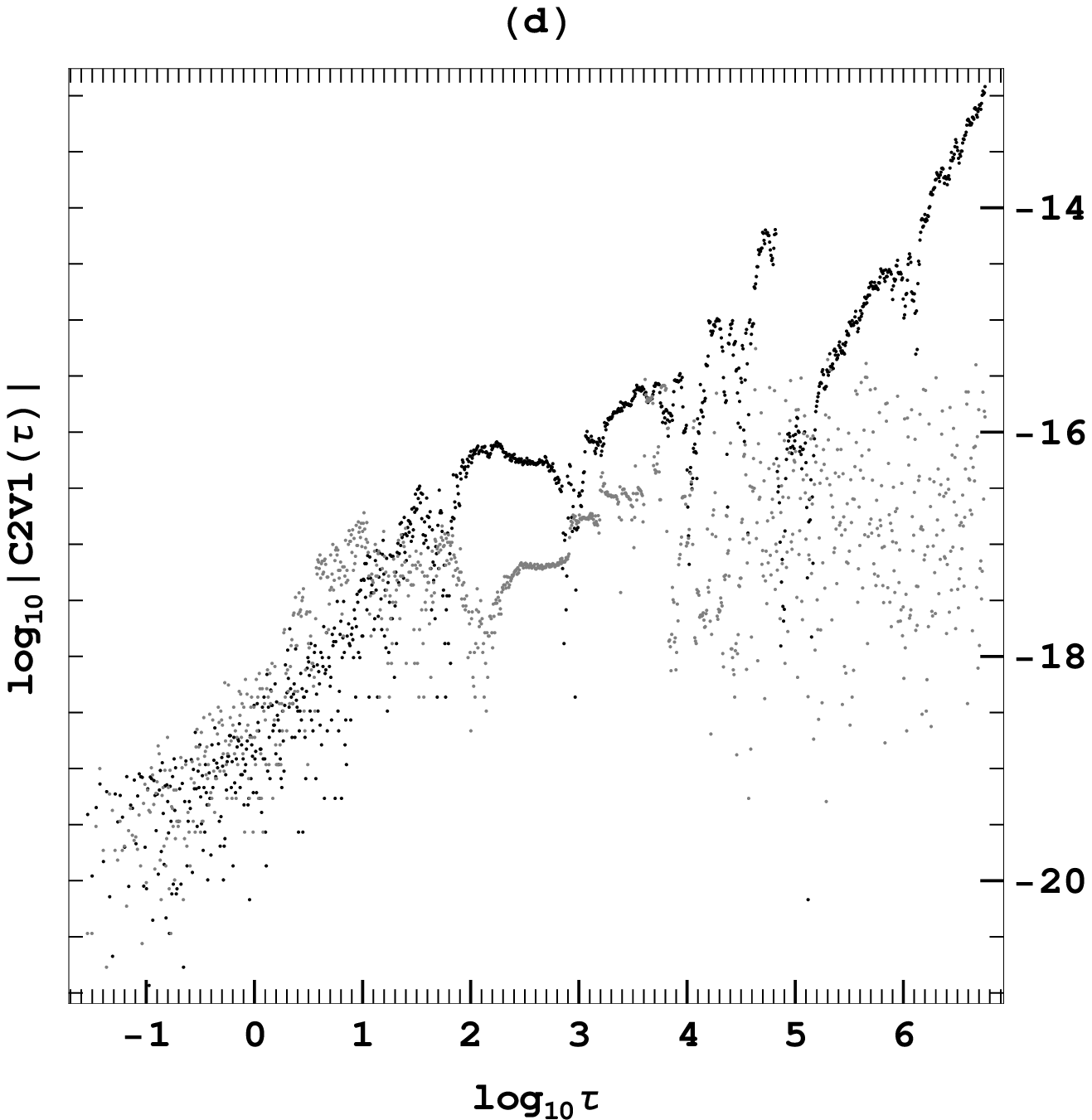}}
 \centerline{\includegraphics[width=0.4 \textwidth] {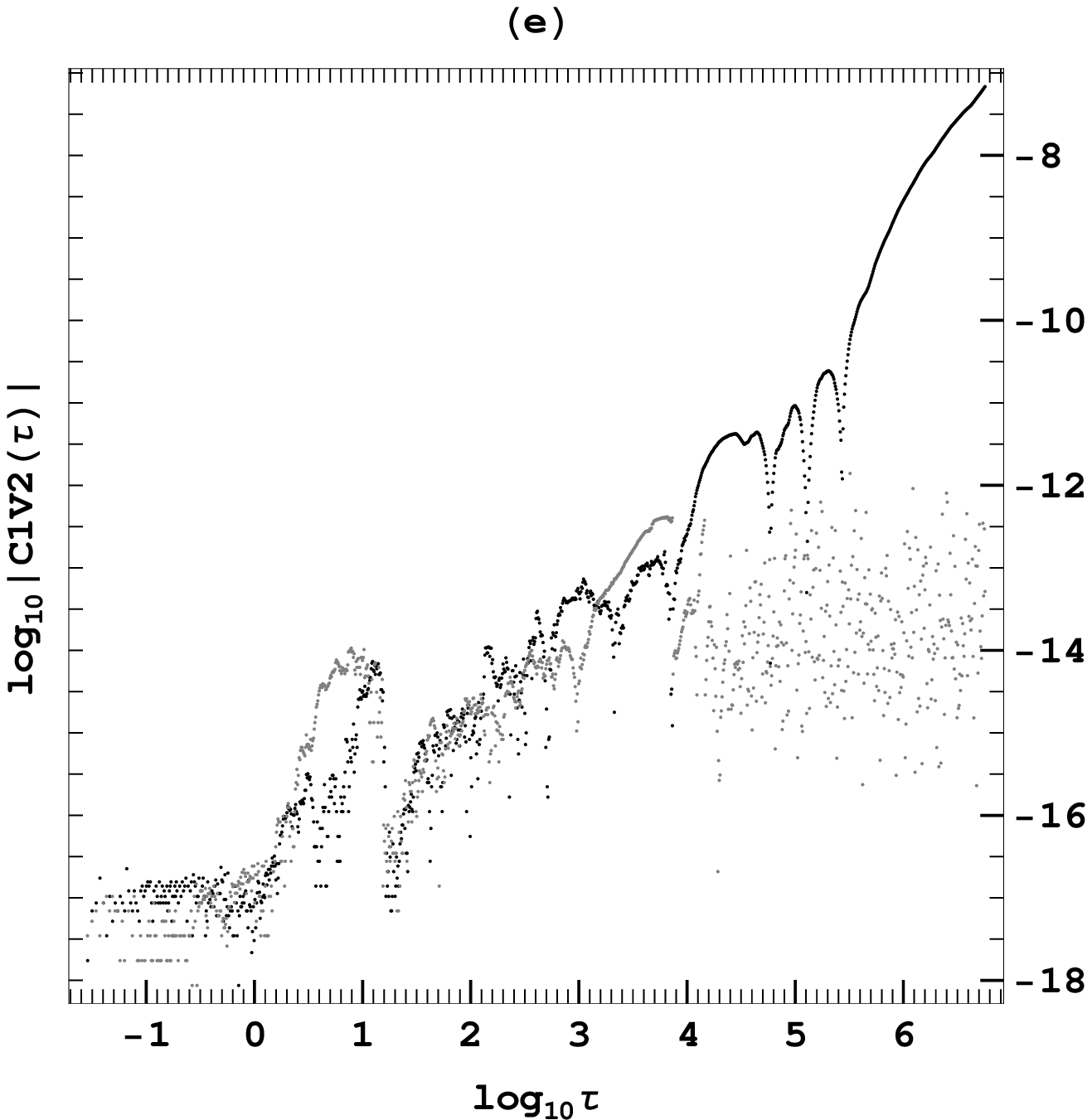}
 \includegraphics[width=0.4 \textwidth] {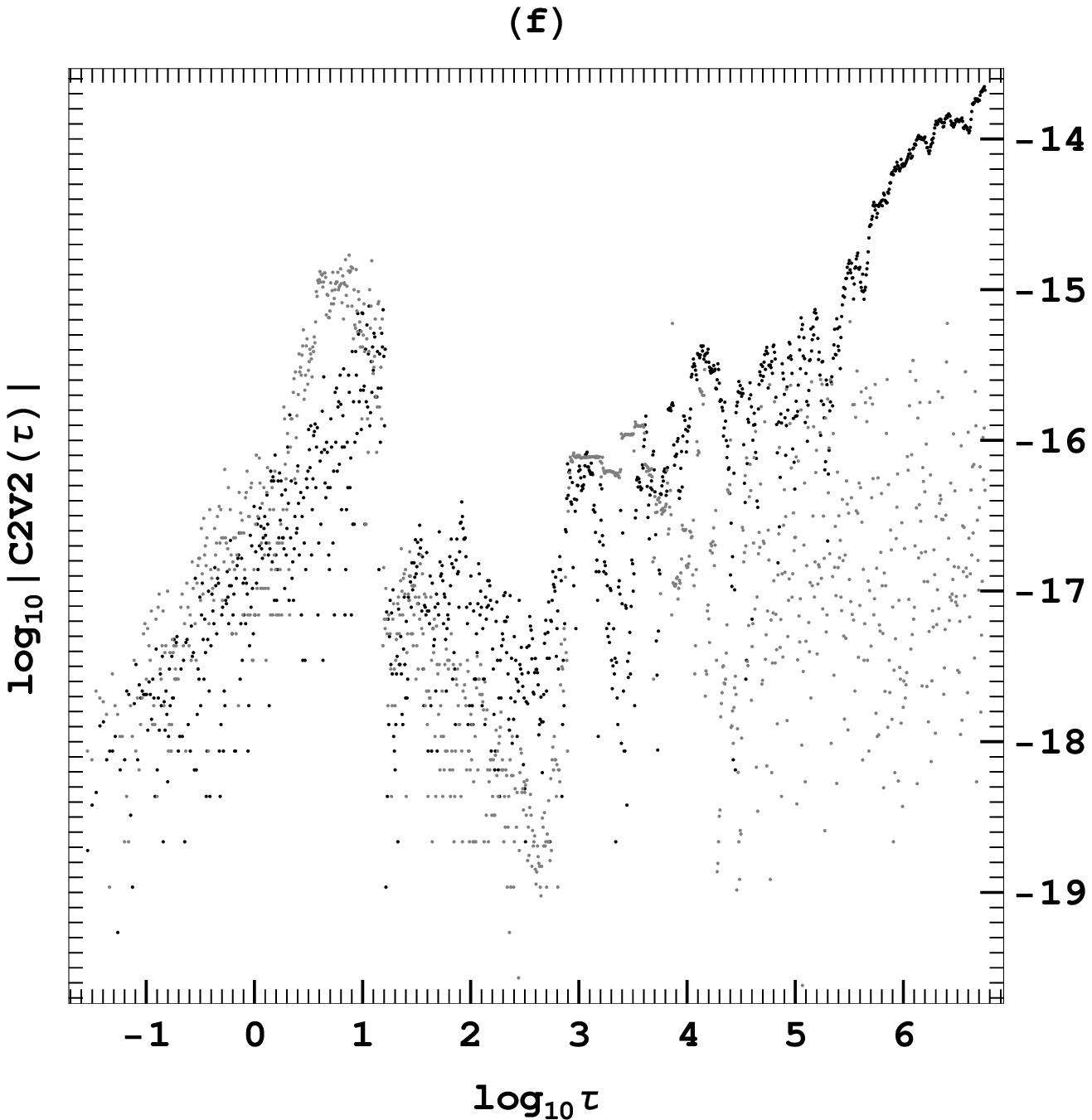}}
 \caption{Accuracy checks of IGEM for the two orbits shown in Fig.~\ref{figEx20}.
 The black points correspond to the regular orbit, while the gray to the chaotic. 
 Panel (a) shows the relative error between two time steps
 $\log_{10}\left|1- \frac{L(\tau)}{L(\tau-d\tau)}\right|$, where the $L(\tau)$ is
 the  value of the Lagrangian function evaluated at time $\tau$. Panel (b) shows
 the overall relative error $\log_{10}\left|1- \frac{L(\tau)}{L}\right|$, where $L$
 is the theoretical value of the Lagrangian function. Panels (c) and (d) show the
 conservation of the constraints $\xi^\alpha \dot{x}_\alpha=0$ (C1V1) and
 $\frac{D \xi^\alpha}{d \tau} \dot{x}_\alpha=0$ (C2V1) (Eq.~(\ref{eq:InDevPre}))
 respectively for the first vector, while panels (e) and (f) show the conservation
 of the constraints $\zeta^\alpha \dot{x}_\alpha=0$ (C1V2) and
 $\frac{D \zeta^\alpha}{d \tau} \dot{x}_\alpha=0$ (C2V2) respectively for the
 second vector. 
}
\label{figEx20Er}
\end{figure*}

 There are three independent and in involution constants of motion, namely, the 
 energy $E$, the $z$ component of the angular momentum $L_z$, and the Lagrangian
 function $L$ itself. The Lagrangian function contains all the variables involved
 in the geodesic motion; thus, it is a very efficient quantity to check the
 accuracy of the numerical scheme applied. There are two relative errors that are
 of interest: first, the relative error between two time steps, i.e.,
 $\log_{10}\left|1- \frac{L(\tau)}{L(\tau-d\tau)}\right|$, and the overall relative    
 error, i.e., $\log_{10}\left|1- \frac{L(\tau)}{L}\right|$, where $L(\tau)$ is the 
 value of the Lagrangian function evaluated at time $\tau$. Fig.~\ref{figEx20Er}(a)
 shows that the relative error between two time steps is of the order of the 
 machine precision; however, the overall relative error seems to grow following a
 power law with exponent $\approx 5/9$ (Fig.~\ref{figEx20Er}(b)). This behavior
 appears to be independent of the character of the orbit; i.e., it does not depend
 on whether the orbit is chaotic (gray points) or regular (black points).

 The deviation vectors $\xi^\alpha$, $\zeta^\alpha$  were set to satisfy the
 constraints (\ref{eq:InDevPre}) during the evolution of the orbits of
 Fig.~\ref{figEx20}. Panels (c)-(d) show how much these constraints were preserved.
 The slopes in these panels again indicate power laws, but each of them has a
 different exponent.

\end{document}